\documentclass[pdflatex, sn-mathphys-num]{sn-jnl}% 

\usepackage{xcolor}
\usepackage{graphicx}%
\usepackage{caption}
\usepackage{capt-of}
\usepackage{subfigure}
\usepackage{multirow}%
\usepackage{amsmath,amssymb,amsfonts}%
\usepackage{amsthm}%
\usepackage{mathrsfs}%
\usepackage[title]{appendix}%
\usepackage{xcolor}%
\usepackage{textcomp}%
\usepackage{manyfoot}%
\usepackage{booktabs}%
\usepackage{algorithm}%
\usepackage{algorithmicx}%
\usepackage{algpseudocode}%
\usepackage{listings}%\usepackage{adjustbox}
\usepackage{rotating}
\usepackage{cellspace}
\usepackage{lmodern}
\usepackage{caption}%
\usepackage{lineno}
\usepackage{hyperref}
\usepackage{orcidlink}

\begin{document}

%\title{First observation of reactor antineutrinos by coherent scattering}
\title{Direct observation of coherent elastic antineutrino-nucleus scattering}

\author[1]{\fnm{N.} \sur{Ackermann}\orcidlink{0009-0000-2122-5186}}

\author[1]{\fnm{H.} \sur{Bonet}}

\author[1,a]{\fnm{A.} \sur{Bonhomme}\orcidlink{0000-0002-0218-2835}}

\author*[1]{\fnm{C.} \sur{Buck}\orcidlink{0000-0002-5751-5289}}\email{christian.buck@mpi-hd.mpg.de}
%\equalcont{These authors contributed equally to this work.}

\author[2]{\fnm{K.} \sur{F\"{u}lber}}

\author[1,b]{\fnm{J.} \sur{Hakenm\"{u}ller}\orcidlink{0000-0003-0470-3320}}

\author[1]{\fnm{J.} \sur{Hempfling}}

\author[1]{\fnm{G.} \sur{Heusser}}

\author[1]{\fnm{M.} \sur{Lindner}\orcidlink{0000-0002-3704-6016}}

\author[1]{\fnm{W.} \sur{Maneschg}\orcidlink{0000-0003-0320-7827}}

\author[1]{\fnm{K.} \sur{Ni}}

\author[3]{\fnm{M.} \sur{Rank}}

\author[1,c]{\fnm{T.} \sur{Rink}\orcidlink{0000-0002-9293-1106}}

\author[1]{\fnm{E.} \sur{S\'{a}nchez Garc\'{i}a}\orcidlink{0000-0001-8014-4079}}

\author[3]{\fnm{I.} \sur{Stalder}}

\author[1]{\fnm{H.} \sur{Strecker}}

\author[2]{\fnm{R.} \sur{Wink}}

\author[3,d]{\fnm{J.} \sur{Woenckhaus}\orcidlink{0009-0003-6621-5979}}

\affil[1]{\orgname{MPIK Heidelberg}, \orgaddress{\street{Saupfercheckweg 1}, \postcode{69117}~\city{Heidelberg}, \country{Germany}}}

\affil[2]{\orgdiv{PreussenElektra GmbH}, \orgname{Kernkraftwerk Brokdorf}, \orgaddress{\street{Osterende}, \postcode{25576}~\city{Brokdorf}, \country{Germany}}}

\affil[3]{\orgname{Kernkraftwerk Leibstadt AG}, \orgaddress{\postcode{5325}~\city{Leibstadt}, \country{Switzerland}}}

\affil[a]{\orgname{{\it Present address:} IPHC, CNRS}, \orgaddress{\city{67037 Strasbourg}, \country{France}}}

\affil[b]{\orgname{{\it Present address:} Duke University}, \orgaddress{\city{Durham, NC 27708}, \country{USA}}}

\affil[c]{\orgname{{\it Present address:} KIT}, \orgaddress{\street{Hermann-von-Helmholtz-Platz 1}, \postcode{76344}~\city{Eggenstein-Leopoldshafen}, \country{Germany}}}

\affil[d]{\orgname{{\it Present address:} PSI}, \orgaddress{\street{Forschungsstrasse 111}, \postcode{5232}~\city{Villigen}, \country{Switzerland}}}

\abstract{Neutrinos are elementary particles that interact only very weakly with matter. Neutrino experiments are, therefore, usually big, with masses in the multi-tonne range. The thresholdless interaction of coherent elastic scattering of neutrinos on atomic nuclei leads to greatly enhanced interaction rates, which allows for much smaller detectors. The study of this process gives insights into physics beyond the Standard Model of particle physics. The \textsc{CONUS+} experiment was designed to first detect elastic neutrino–nucleus scattering in the fully coherent regime with low-energy neutrinos produced in nuclear reactors. For this purpose, semiconductor detectors based on high-purity germanium crystals with extremely low-energy thresholds were developed. Here we report the first observation of a neutrino signal with a statistical significance of 3.7\,$\sigma$ from the \textsc{CONUS+} experiment, operated at the nuclear power plant in Leibstadt, Switzerland. In 119~days of reactor operation ($395\pm106$) neutrinos were measured compared with a predicted number from calculations assuming Standard Model physics of ($347\pm59$) events. With increased precision, there is potential for fundamental discoveries in the future. The CONUS+ results in combination with other
measurements of this interaction channel might therefore mark a starting point for a new era in neutrino physics.}

\maketitle

%\linenumbers

\section*{Introduction}\label{sec1}
Neutrinos are known for their tiny interaction rate with matter. This is why they are usually very hard to detect, despite their high abundance. The most common detection channels such as the inverse beta decay reaction or neutrino-electron scattering usually require target masses in the ton to kiloton scale. In the Standard Model (SM) of particle physics, neutrinos can couple to quarks by the exchange of a mediating Z boson. For small momentum exchanges, the possibility of coherent scattering of neutrinos on the sum of all nucleons of an atomic nucleus was predicted in 1974~\cite{Freedman:1973yd}. For this interaction the reaction rate (cross-section) is enhanced by few orders of magnitude as it scales approximately with the squared number of neutrons in the target nucleus. Therefore, it is in principle possible to construct neutrino detectors on the kilogram scale using this channel. 

It took 43 years after its prediction until CE$\nu$NS was first detected in 2017 by the COHERENT experiment in a scintillating crystal of cesium iodide~\cite{Coherent:2017}. Later, COHERENT confirmed the measurement with argon~\cite{COHERENT:2020iec} and germanium~\cite{COHERENT:2025vuz} as target materials. Here, the neutrinos are generated at a Spallation Neutron Source (SNS) when pions decay at rest. A complementary approach for CE$\nu$NS detection is to use nuclear reactors as a source. An advantage is the lower neutrino energies as compared to the SNS source, which offers an enhanced sensitivity for several parameters in beyond the standard model (BSM) theories. At higher neutrino energies above approximately 10\,MeV, there is a transition from the fully to the partially coherent regime, and uncertainties related to the nuclear structure become relevant. Measurements at reactor can be used to get a clean coherent scattering signal and set an anchor point for the cross-section. With this knowledge, experiments such as COHERENT can study the nuclear structure of the corresponding target isotope with higher-energy neutrinos. Nuclear reactors provide a very high flux of pure electron antineutrino, whereas the accelerator-based SNS source involves several types of neutrinos, including muon neutrinos and antineutrinos.

There is a long history of successful experiments using reactor antineutrinos as sources, including the first neutrino detection in 1956~\cite{Cowan:1956rrn}. In recent years, reactor experiments allowed to study neutrino oscillation parameters~\cite{KamLAND:2008dgz, DayaBay:2022orm, DoubleChooz:2019qbj, RENO:2018dro} and constrain the existence of sterile neutrinos~\cite{STEREO:2022nzk, PROSPECT:2020sxr, DANSS:2018fnn, RENO:2020hva}. Today, there is a worldwide effort to measure CE$\nu$NS near nuclear reactors~\cite{CONNIE:2021ggh, nGeN:2022uje, NUCLEUS:2019igx, Colaresi:2022obx, CONUS:2020skt, Ricochet:2021rjo, Kerman:2016jqp, NEON:2022hbk, Yang:2024exl, RED-100:2024izi}. In terms of neutrino oscillation studies, it is a complementary approach, since the CE$\nu$NS process is sensitive to the three known neutrino flavors, whereas the inverse beta decay as a standard detection technique is only sensitive to electron antineutrinos. There is a wide range of studies that can be addressed in CE$\nu$NS measurements. For example, they are highly relevant for probing nuclear structures, for astrophysical studies, and for present and future dark matter experiments, which are limited in sensitivity by the CE$\nu$NS rate of solar neutrinos. Recently, indications for a CE$\nu$NS signal were found in dark matter experiments with xenon as target~\cite{XENON:2024ijk, PandaX:2024muv}. Moreover, \textsc{Conus+} technology has the potential to use neutrinos for reactor monitoring and safeguard applications in the future.

\section*{The CONUS and CONUS+ experiments}\label{sec2}
The \textsc{Conus} experiment~\cite{CONUS:2020skt} began operating in 2018 at the nuclear power plant in Brokdorf (KBR), Germany, and took data until 2022. The experimental setup consisted of four high-purity germanium (HPGe) detectors. Each diode had a mass of about 1\,kg leading to a total fiducial germanium mass of $3.73\pm0.02$~kg~\cite{Bonet:2020ntx}. As a final result of the measurement at KBR, the neutrino flux was constrained with 90\% confidence level to a factor 1.6 above the signal expectation~\cite{CONUSCollaboration:2024oks}. This world-best upper limit on the CE$\nu$NS interaction rate at nuclear reactors so far allowed us to exclude significant deviations from the SM or to test the standard description of signal quenching due to dissipation effects in the germanium material in the energy region of interest.

In 2023, the \textsc{Conus} setup moved to another power plant in Leibstadt, Switzerland (KKL), since the reactor at KBR stopped operation. At KKL the experiment continued as \textsc{Conus+}. Here, antineutrinos are created in a boiling water reactor with a thermal power of 3.6\,GW. The setup is placed at a distance of about 20.7\,m from the center of the KKL reactor core. Before installation, the HPGe detectors called C$2-$C5 were refurbished to improve the energy threshold and the detection efficiency at low energy. In this way, the sensitivity was improved despite a higher level of environmental radioactivity and a slightly lower nominal neutrino flux of $1.5\cdot$10$^{13}$~antineutrinos/(cm$^{2}$s) at the new site. The predicted rate of neutrino interactions increased almost an order of magnitude, mainly due to the improved energy threshold and trigger efficiency~\cite{CONUS:2024lnu}. 

\begin{figure}[ht]
    \centering
    \includegraphics[width=\textwidth]{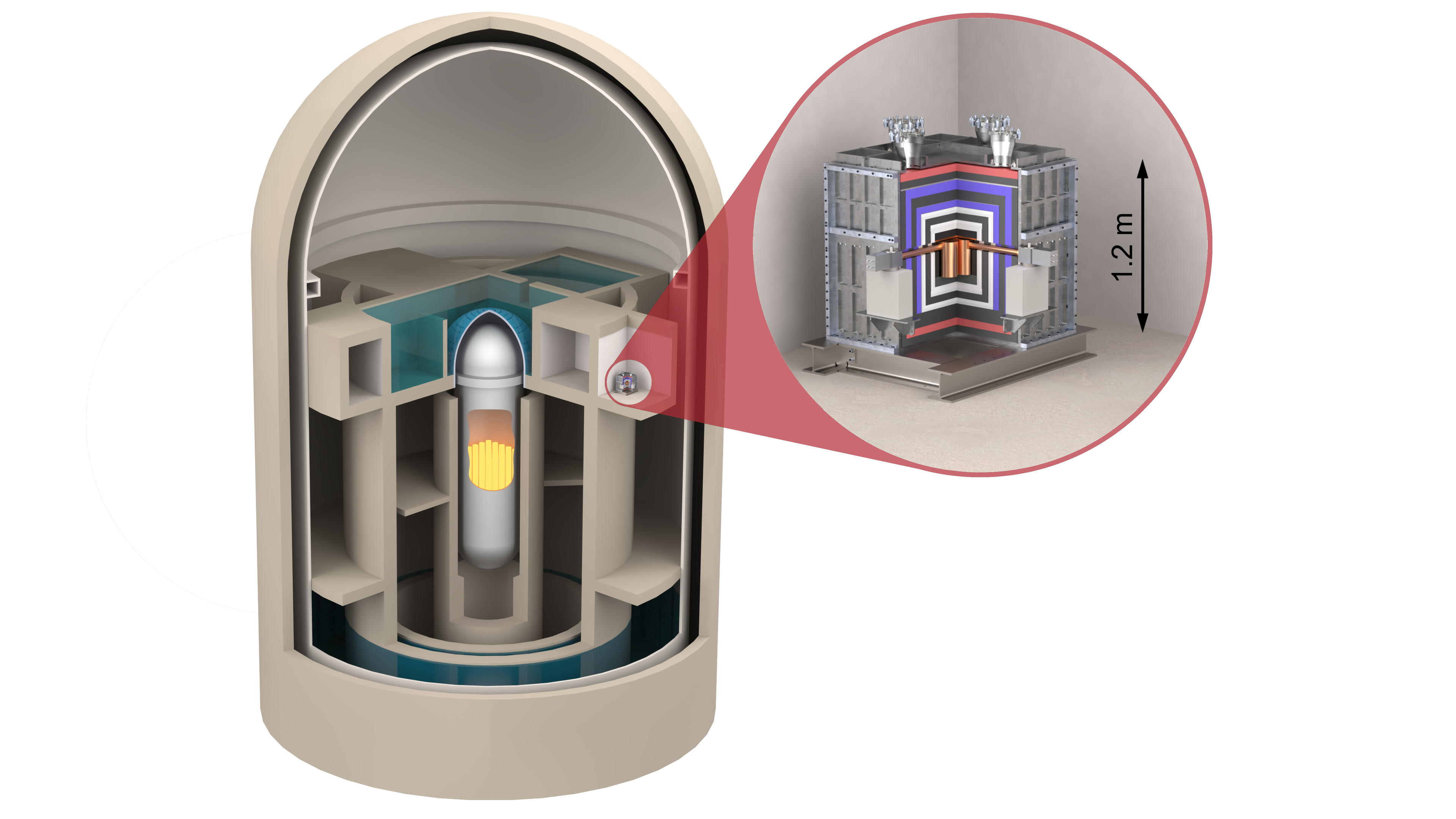}
\caption{\textbf{Configuration of the experimental setup.} The \textsc{Conus+} shield with a weight of about 10~tons is installed inside the reactor building of the nuclear power plant in Leibstadt at a distance of 20.7\,m from the reactor core. It is mechanically stabilized by a stainless steel (silver) frame to assure integrity of the setup in case of an earthquake. Layers of lead (black) reduce the impact from external gamma radiation, polyethylene (red) and boron-doped polyethylene (white) moderate and capture neutrons emitted from the reactor. Two layers of plastic scintillator plates are equipped with photo multipliers (blue) and are used as an active veto system discriminating background from cosmic muons. In the central detector chamber, four germanium diodes are operated inside radiopure copper cryostats and connected to electrically powered cryocoolers.}
\label{conus_shield}    
\end{figure}

The search for CE$\nu$NS in nuclear reactors is a challenging task for various reasons. Since a high neutrino flux is required, positions close to the reactor core inside the reactor building are preferred. The environment inside this inner control zone is quite different from the working conditions in a common research laboratory. There are several constraints related to material restrictions, earthquake safety, access and data transfer. Appropriate solutions on all these topics were found in close cooperation with the KBR and KKL staff. In addition, there is limited protection against cosmic radiation, as the overburden at the \textsc{CONUS+} site corresponds only to the equivalent 7.4\,m water. In general, the radioactivity level has to be kept under control to perform successful rare event searches. For example, cosmic muons produce electromagnetic cascades and neutrons in the building structure and shield materials. These cascades can create event signatures similar to neutrinos in the HPGe detectors. Mitigation of detector signals created by such cosmic radiation or environmental radioactivity (background events) is achieved by using an effective shield structure around the detectors~\cite{CONUS:2024lnu} as depicted in Fig.\ref{conus_shield}. Furthermore, the energy of the nuclear recoils after neutrino scattering is very low. The unit used for the energy measured by the Ge detectors is given in eV ($1.6\cdot10^{-19}$\,J) and should be interpreted as ionization energy. It was a longstanding effort to reach the required threshold levels in the detectors. Our currently lowest threshold level of 160\,eV is only two orders of magnitude above the typical semiconductor band gap, which defines the minimum energy to create one electron-hole pair.

\section*{Neutrino signal observation in CONUS+}\label{sec3}

The basic concept of antineutrino detection in \textsc{Conus+} is to measure an energy spectrum under stable conditions in phases for which the reactor is running or stopped (reactor on and off). Most events originate from cosmic radiation, which is fully independent of the reactor operation condition. During on-phases additional neutrino events are expected with a characteristic spectral shape. The reactor off-phases without neutrino signal are obtained in maintenance periods for refueling, which are typically once per year with a duration of approximately 1\,month. 
    
\begin{figure}[ht]
\begin{subfigure}
    \centering
    \textbf{a}
    \includegraphics[width=0.47\textwidth]{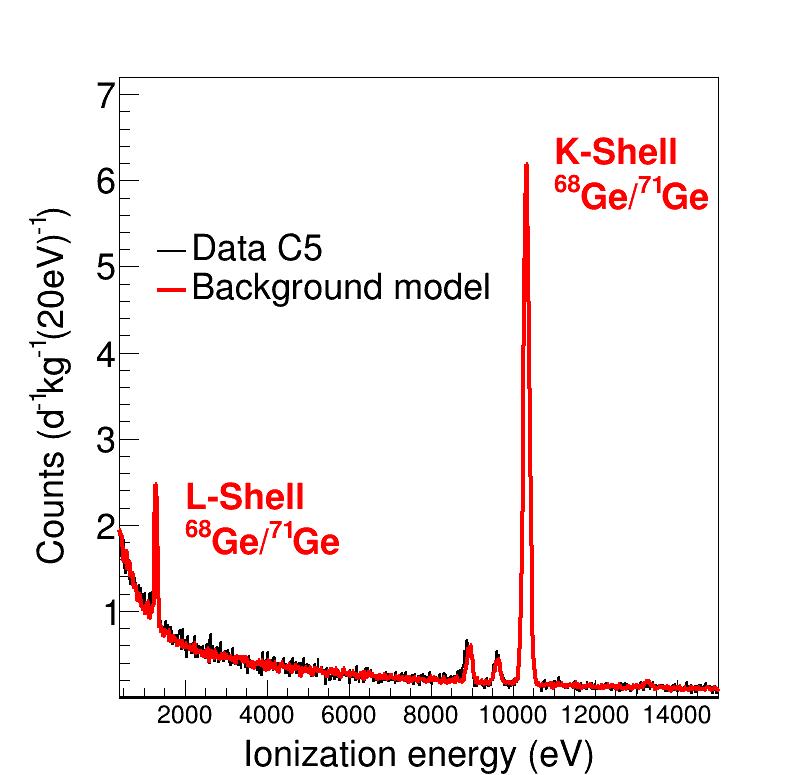}
\end{subfigure}
\begin{subfigure}
    \centering
    \textbf{b}
    \includegraphics[width=0.47\textwidth]{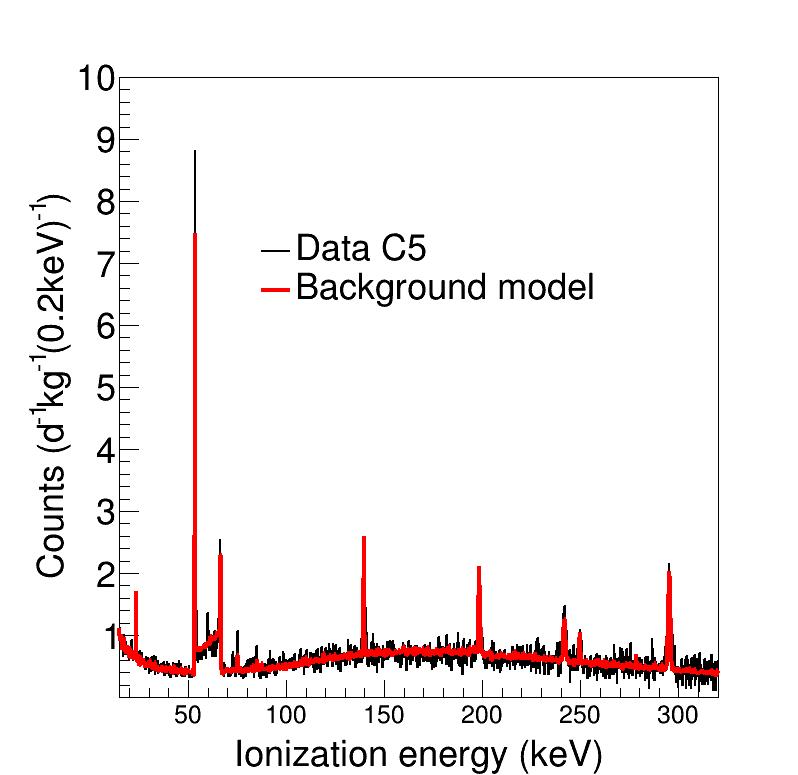}
\end{subfigure}
\caption{\textbf{Comparison of data and background model. a)} The plot shows the background spectrum from 0.4\,keV, directly above the ionization energy range in which the neutrino signal is expected. The data (black line) is shown for the reactor on period and is based on the measurement of the detector with the lowest background in the \textsc{Conus+} analysis (C5). This data is compared to the background model (red line) and found to be in good agreement. \textbf{b)} The high energy channel up to few hundred keV is shown.}
\label{background2}
\end{figure}

The energy spectrum during reactor on period is depicted in Fig.\ref{background2}, together with the expected background calculated using a well-validated GEANT4 framework~\cite{Bonet:2021wjw}. A good understanding of the background composition is mandatory for our neutrino analysis. Following the experience gained at KBR, the background spectra were adjusted to the new KKL site. The contributions of muon-induced, neutron and gamma components were validated in a dedicated measurement campaign with multiple detector technologies before installing the \textsc{Conus+} setup~\cite{CONUS:2024vyx}. Uncorrelated background contributions, which are independent of the reactor thermal power condition, are measured in the reactor off-phase.

The dominant background contributions, such as the muon-induced or neutron background, are identical for the three detectors used in the analysis. In addition to the neutrons, which are created by muons in the materials of the \textsc{Conus+} shield, there is a relevant fraction of direct neutrons up to 100\,MeV energy already produced in the atmosphere in cosmic ray air showers. The background during reactor off was lower since a vessel lid with a thickness of a few cm of steel was placed right above the \textsc{Conus+} room during reactor outage providing additional overburden. According to simulations, this reduction was 3\% for muons and 19\% for direct cosmic neutrons. Other components contributing to the background spectrum are from metastable states in germanium and the radioactive noble gas radon. Variations in the radon level inside the detector chamber were corrected. Otherwise, no significant differences are observed during reactor on and off. The reactor correlated background events, in particular the ones from neutrons produced in the reactor, are thoroughly studied in~\cite{CONUS:2024vyx} and found to be negligible.

The calculation of the signal prediction is based on a method proposed in~\cite{DayaBay:2021dqj}. The neutrino flux depends on the fission fractions of the uranium and plutonium isotopes in the reactor: $^{235}$U, $^{238}$U, $^{239}$Pu and $^{241}$Pu, respectively. The average contribution of these isotopes to the flux during data collection with reactor on is 53\%, 8\%, 32\% and 7\%. In principle, there is no energy threshold for the CE$\nu$NS interaction itself and therefore there is a potential to measure neutrinos even below the threshold for the inverse beta decay reaction of about 1.8\,MeV. Beyond the reactor conditions and its thermal power, the neutrino rate measured in the detector strongly depends on dissipation processes in the germanium crystals. The ionization energy observed in the detector is reduced with respect to the deposited recoil energy, a characteristic known as quenching. There was a debate, if this quenching factor is enhanced at low energies as compared to the Lindhard theory~\cite{lindhard1963range, Collar:2021fcl}. From \textsc{Conus} data, there is no indication of any deviation from the Lindhard model~\cite{Bonhomme:2022lcz, CONUSCollaboration:2024oks}.  

A significant contribution to the systematic uncertainty is related to the precision of the energy scale calibration. At very low energies close to the energy region of interest, X-rays emitted in radioactive decays inside the detector crystals are used for calibration purposes. There are prominent lines around 10.4\,keV corresponding to the binding energies of the K-shells and around 1.3\,keV from L-shells of Ge isotopes, as seen in the left plot of Fig.\ref{background2}. Toward the end of the first data collection period in \textsc{Conus+}, the detectors were irradiated with neutrons from a californium source outside the shield to increase the statistics in the corresponding lines and reduce the uncertainty of the energy scale to less than 5\,eV. Small energy non-linearity effects close to the detection threshold induced by the data acquisition system~\cite{Bonhomme:2022lcz} were measured with a pulse generator and corrected accordingly. 

The energy threshold of the detectors is estimated for each of them individually~\cite{CONUS:2024lnu}. The lowest value for the C3 detector is only 160\,eV. For the other two detectors used in the analysis, the thresholds after energy non-linearity correction were set slightly above, at 170\,eV and 180\,eV. One of the four detectors (C4) showed significant instabilities in the rate and was therefore removed from the data set. The thresholds were defined in a way to assure that contributions from electronic noise and microphonics are negligible in the region of interest. The trigger efficiency was determined for each detector using a pulse generator and was found to be close to 100\% above the thresholds.  

Another crucial requirement for detecting CE$\nu$NS at a nuclear reactor is the stability of environmental parameters, electronic noise and background rates. For example, temperature fluctuations can induce cryocooler power variations, which might create microphonic events. Microphonic noise and rate correlations with room temperature were further reduced compared to previous analyses~\cite{CONUS:2020skt, CONUSCollaboration:2024oks} by an improved cooling system~\cite{CONUS:2024lnu}. The stability of the detector parameters such as energy resolution or trigger efficiency was regularly checked with the pulse generator. Variations of the noise peak were carefully monitored and data were only selected for the analysis in case they were below a defined level. 

The data set used in the analysis reported here includes reactor on periods between November 2023 and July 2024 (327~kg\,d) and an off period during reactor outage in May 2024 (60~kg\,d). The data sets of the three detectors are fitted simultaneously in an energy window between 160\,eV and 800\,eV. The signal is extracted based on a profile likelihood ratio test. Systematic uncertainties are considered with Gaussian pull terms. The data acquisition system in principle allows for background rejection by studying the shape of digitized pulses~\cite{Bonet:2023kob}. This option was not yet applied in the current analysis, but it is planned to be used in future analyses.

\begin{figure}
\begin{subfigure}
    \centering
    \textbf{a}
    \includegraphics[width=0.95\textwidth]{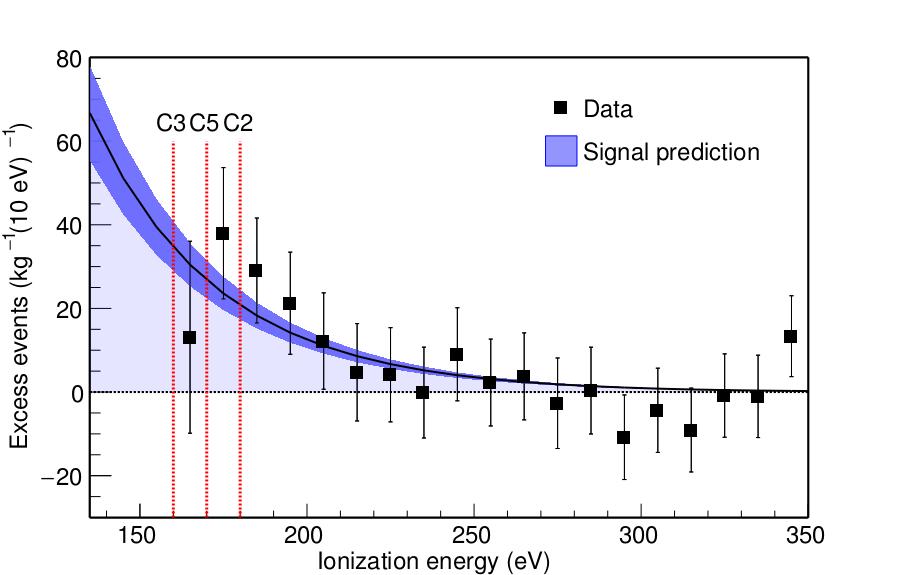}
\end{subfigure}
\begin{subfigure}
    \centering
    \textbf{b}
    \includegraphics[width=0.46\textwidth]{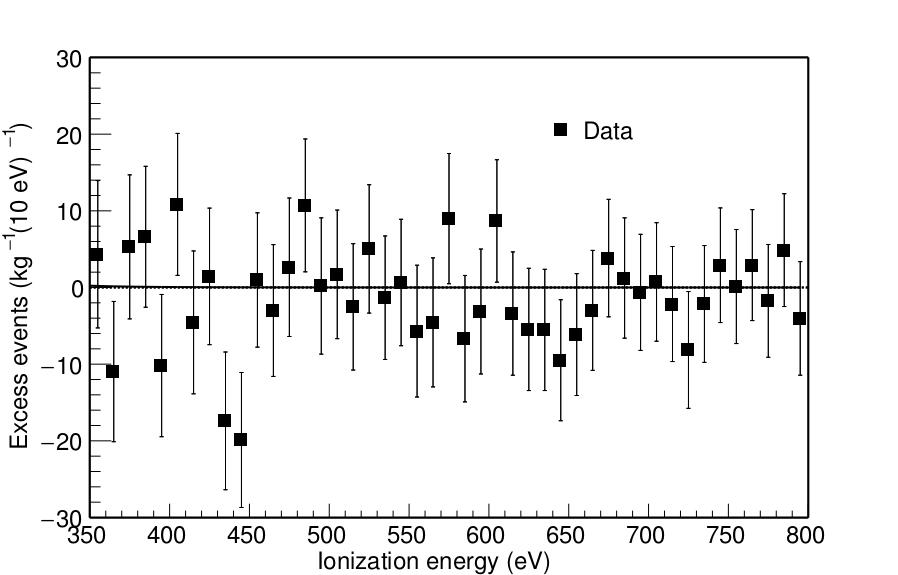}
\end{subfigure}
    \centering
    \textbf{c}
    \includegraphics[width=0.41\textwidth]{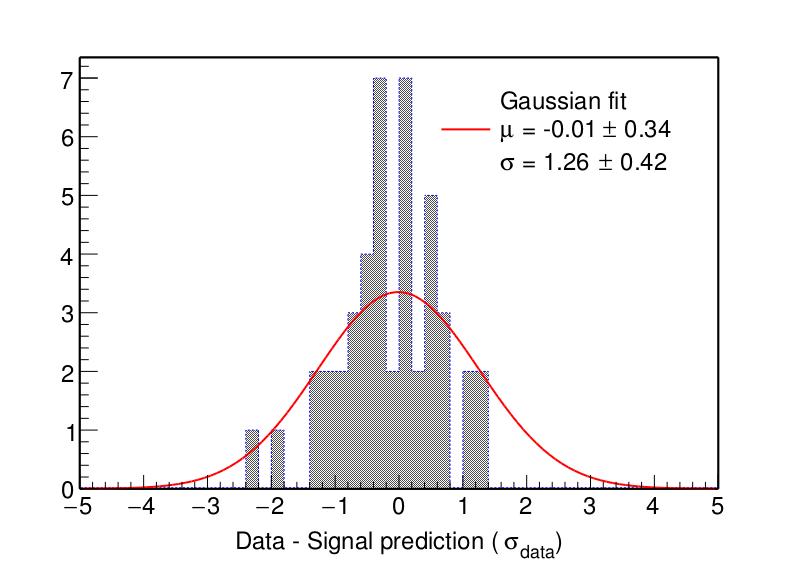}
\caption{\textbf{Neutrino signal. a)} The difference between data in the full reactor on time and the background model scaled to the total detector mass is shown. The vertical error bars represent the combined statistical uncertainties of data points and background model at 68\% CL (1$\sigma$). At low energies the rise from the neutrino signal can be seen. The line shows the predicted signal shape for comparison including a 68\% CL uncertainty band. The red vertical lines indicate the energy thresholds of the three detectors used in the analysis. In the first bin only C3 contributes, in the second C3+C5 and above 180\,eV all three detectors. Residual systematic effects from energy non-linearity, background model uncertainties and the fit method are not incorporated in this plot. \textbf{b)} This graph illustrates the good agreement between the data and the background model above the signal region from $350-800$\,eV. \textbf{c)} This histogram shows the spread of the data points around the background model for the same energy interval as in b).}
\label{result}
\end{figure}

From the combined fit, a neutrino signal of ($395\pm106$) events in the reactor on data set was obtained. The significance of this event excess corresponds to $3.7\,\sigma$. The neutrino signal at low energies of the spectrum is illustrated in Fig.\ref{result}. Fits using just single detectors independently give consistent results. The result was cross-checked in two implementations of the likelihood fit. Additional systematic uncertainties related to background model, non-linearity correction and fit systematics were studied independently and are included in the final result.  

\section*{Impact and outlook}\label{sec4}
The \textsc{Conus+} signal of ($395\pm106$) measured neutrinos is fully consistent with the expectation of ($347\pm59$) events. This implies agreement of the \textsc{Conus+} data with the CE$\nu$NS cross-section of the SM and the estimated antineutrino flux based on the thermal power of the reactor. Moreover, the detected rate is in very good agreement with the predicted Ge quenching using the Lindhard theory with a quenching parameter as measured in~\cite{Bonhomme:2022lcz}. 
The deviations from Lindhard theory claimed in~\cite{Collar:2021fcl} and the excess reported in~\cite{Colaresi:2022obx}, which is based on it, are both ruled out by this result. The claim was already disfavored before by the independent measurement of the quenching factor~\cite{Bonhomme:2022lcz}, previous \textsc{Conus} results~\cite{CONUSCollaboration:2024oks} and theoretical considerations~\cite{AtzoriCorona:2023ais}.

In Table~\ref{tab2}, the ratio of measured and predicted neutrino interactions is shown for the \textsc{Conus+} result. This ratio is compared to the ratios measured in the COHERENT measurements at higher neutrino energies using multiple target nuclei and the results from coherent scattering of solar neutrinos. The use of different target nuclei allows to study the quadratic enhancement of the cross-section by the number of neutrons in the target. In the COHERENT data for Ge, the measured rate was found slightly below the predicted value, although not highly significant. Such a deficit was not confirmed in the \textsc{Conus+} data. The combination of the CE$\nu$NS result of \textsc{Conus+} with results of the germanium target data of the COHERENT experiment can in principle be used to extract information about nuclear form factors in neutrino light.  

\begin{table}
\caption{Comparison with other CE$\nu$NS measurements.}\label{tab2}%
\setlength\extrarowheight{3pt}
\begin{tabular}{@{}lccccc@{}}
\toprule
Source & Target  & $\nu$ energy [MeV] & flux [cm$^{-2}$s$^{-1}$] &data & data/SM prediction  \\
\midrule
Accelerator~\cite{COHERENT:2021xmm}    & Cs   & $\sim10-50$  &  $5\cdot$10$^{7}$ & $306\pm20$&  $0.90\pm0.15$ \\
Accelerator~\cite{COHERENT:2020iec}    & Ar   & $\sim10-50$  & $2\cdot$10$^{7}$ & $140\pm40$&  $1.22\pm0.37$\\
Accelerator~\cite{COHERENT:2025vuz}    & Ge   & $\sim10-50$  & $5\cdot$10$^{7}$ & $21\pm7$& $0.59\pm0.21$ \\
Sun~\cite{XENON:2024ijk}   & Xe   & $< 15$  & $5\cdot$10$^{6}$ & $11\pm4$ &  $0.90\pm0.45$ \\
Sun~\cite{PandaX:2024muv}  & Xe   & $< 15$  & $5\cdot$10$^{6}$ & $4\pm1$ &  $1.25\pm0.52$ \\
Reactor    &  Ge  & $< 10$  & $1.5\cdot$10$^{13}$ & $395\pm106$& $1.14\pm0.36$\\
\botrule
\end{tabular}
\end{table}

After the first detection of CE$\nu$NS at a nuclear reactor, as reported in this article, the next step will be a more precise measurement of the CE$\nu$NS cross-section. Higher precision can be achieved by increasing the target mass, lowering the energy threshold of the detectors and longer operation times, in particular in the reactor off phases. Therefore, 3 of the 4 \textsc{Conus+} detectors were replaced in November 2024 by a newer generation. These new detectors have a larger mass of 2.4\,kg each. First characterizations in the MPIK laboratory indicated even lower energy thresholds. The detector with the lowest energy threshold in the first \textsc{Conus+} run (C3) was kept as a reference for a better comparison between the phases of the experiment. With this detector configuration it is planned to measure for another few years.

A high-statistics CE$\nu$NS measurement might open a new phase in fundamental physics and will allow to study physics within and beyond the SM~\cite{Lindner:2024eng}. The measured CE$\nu$NS rate can, for example, be affected by new mediator particles similar to the Z boson, electromagnetic properties of neutrinos, or non-standard interactions. Moreover, it is possible to study the Weinberg angle at low energies, the existence of sterile neutrinos, or supernovae astrophysics. With a precise CE$\nu$NS measurement, more can be learned about neutrino sources as the Sun or nuclear reactors. The evolution of reactor thermal power and fissile isotope concentrations in fuel elements could be monitored with rather small and mobile neutrino detectors. In summary, there is a wide range of topics ranging from BSM theories, nuclear physics, and astrophysics that can be addressed with CE$\nu$NS measurements at nuclear reactors. The \textsc{Conus} and \textsc{Conus+} experiments are pioneering in this field.

\section*{Acknowledgments}
We thank all divisions and workshops involved at
the Max-Planck-Institut für Kernphysik in Heidelberg to set up the \textsc{CONUS+} experiment, in particular T.~Apfel, M.~Reissfelder, T.~Frydlewicz and J.~Schreiner. The authors also thank Mirion Technologies (Canberra) in Lingolsheim for the detector upgrades and their highly professional support. Our deepest gratitude goes to Kernkraftwerk Leibstadt AG for hosting and supporting the \textsc{CONUS+} experiment with special thanks to P.~Graf, P.~Kaiser, L.~Baumann, R.~Meili and A.~Ritter. All authors are members of the \textsc{CONUS} Collaboration.

\newpage 

%\setcounter{figure}{0}
%\renewcommand{\figurename}{Extended Data Fig.}
%\setcounter{table}{0}
%\renewcommand{\tablename}{Extended Data Table}

%\clearpage

\section*{Supplemental material}\label{secA1}

\subsection*{Data taking}\label{secA1.1}

Data from the \textsc{Conus+} run~1 analysis presented in this publication correspond to the period from November 2023 to July 2024. Three of the four detectors, named C2, C3, and C5, are considered for analysis with a total active mass of 2.83$\pm$0.02~kg~\cite{Bonet:2020ntx}. After selection cuts to remove time periods with unstable noise conditions, deficient radon flushing, and a few days with a strong contribution of microphonic events, the exposure considering the active mass is 327~kg~days with reactor on and 60~kg~days with the reactor off.

\begin{sidewaysfigure*}
	\includegraphics[width=0.3\textwidth]{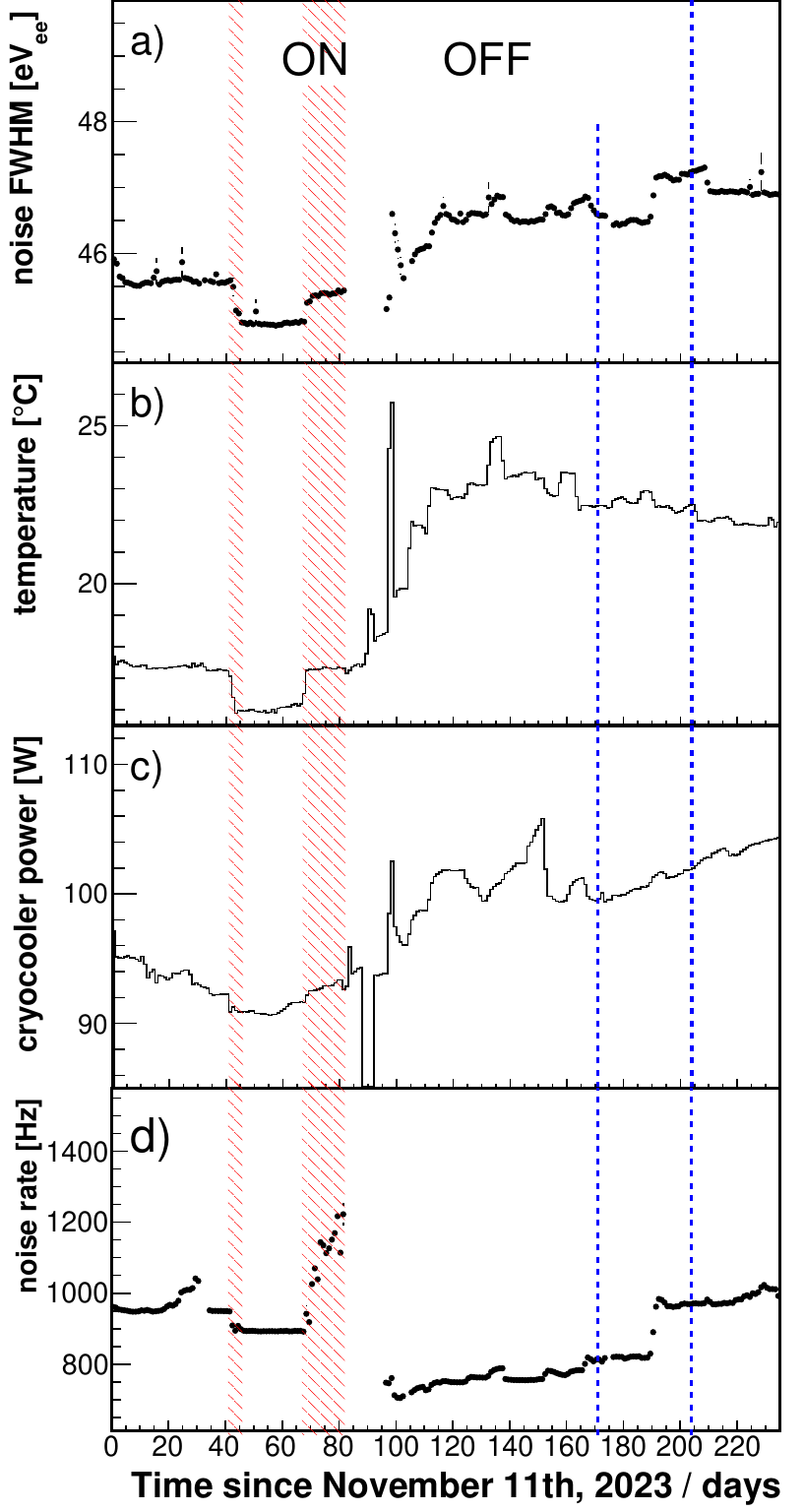}
 \hspace{1.0cm}
 	\includegraphics[width=0.3\textwidth]{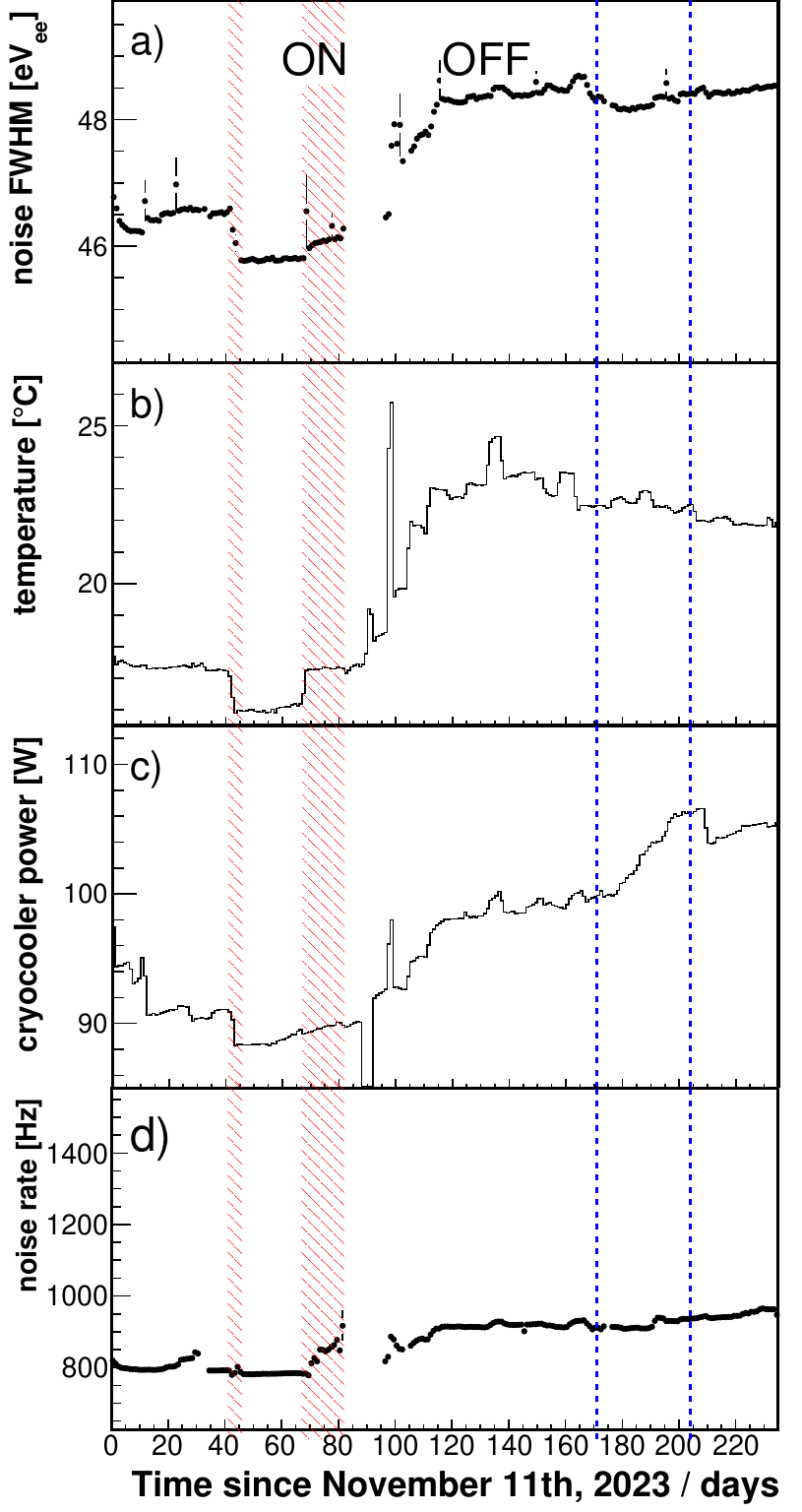}
     \hspace{1.0cm}
    \includegraphics[width=0.3\textwidth]{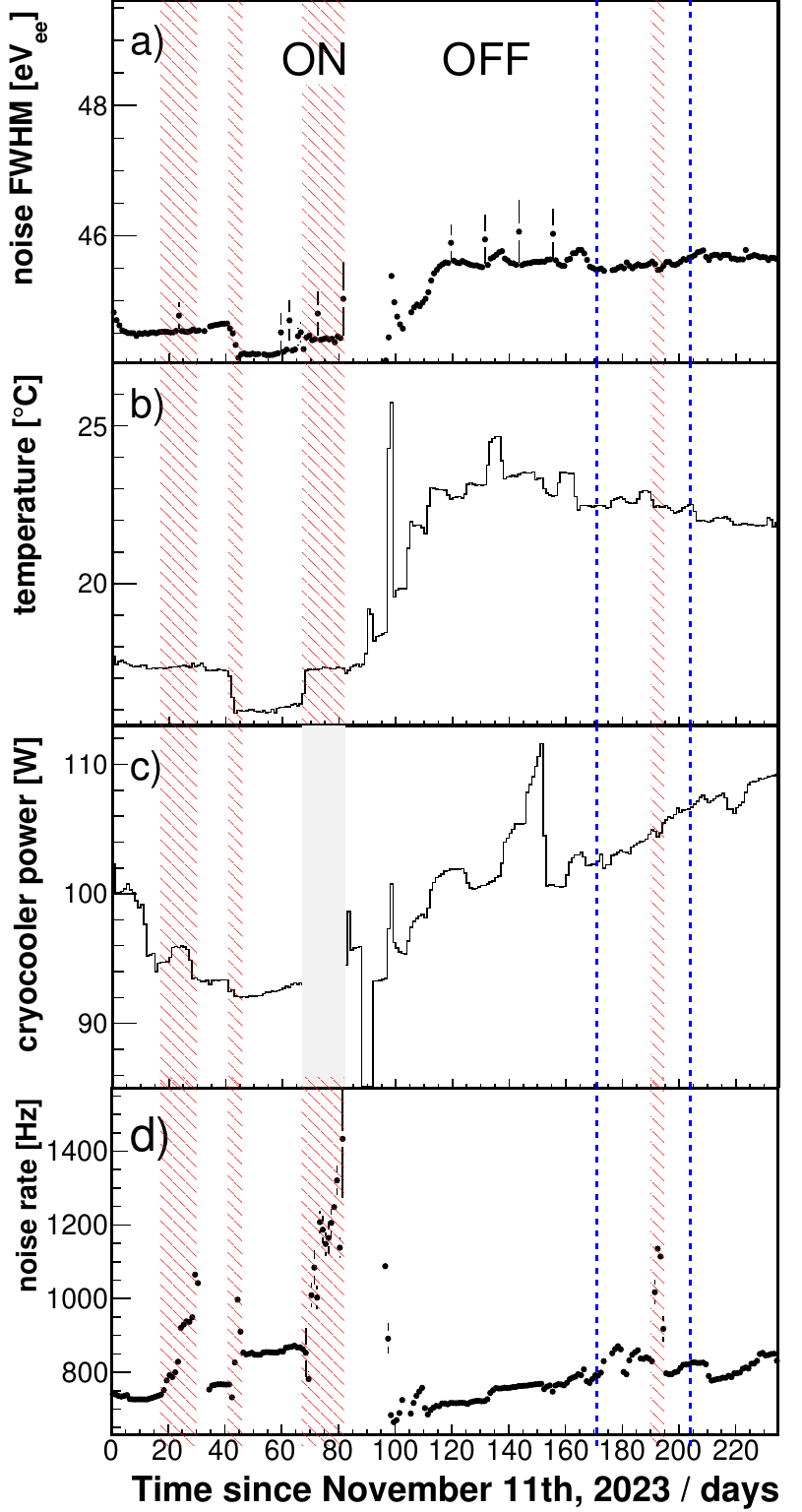}
    \captionof{figure}{\textbf{Stability.} Evolution of parameters for C5 (left), C2 (center), and C4 (right) detectors during run~1. From top to bottom: a) FWHM of noise peak, b) air temperature close to the detector preamplifier, c) cryocooler power consumption, and d) event trigger rate. Red areas were excluded from the analysis, and blue dashed lines indicate the period of reactor outage.
    \label{fig:methods_1}}
\end{sidewaysfigure*}

The evolution of the main environmental parameters during reactor on and off is shown in Extended Data Fig.~\ref{fig:methods_1} for the three detectors. The shape of the noise peak reconstructed in the lowest channels of the data acquisition system (DAQ) was found to follow a Gaussian distribution~\cite{CONUS:2024lnu} and its width (FWHM) was monitored over time with variations below 2\,eV$_{ee}$ (electron equivalent energy). The data set was selected to ensure that the noise rate variations are below 20\%. 

In a dedicated study, we confirmed the correlation between noise rate and cryocooler power. Stability was improved compared to the CONUS setup in Brokdorf (KBR) by replacing the 2-fan ventilation system with a water-cooled chiller system for the pulse tube cryocoolers~\cite{CONUS:2024lnu}. Even at cryocooler power variations of up to 30\,W no correlation was observed with the count rate in the region of interest above 160\,eV$_{ee}$.

\subsection*{Detector response and energy scale}\label{secA1.2}

The trigger efficiency was determined by injecting artificial signals produced by a pulse generator with the same rise time as the physical signals. The pulses are injected through a specific circuit implemented in the HPGe preamplifier. A detailed scan allowed us to measure the detector response as a function of the energy.~\cite{CONUS:2024lnu} 
The trigger efficiency remains over 90\% down to 140~eV$_{ee}$ for all detectors. The evolution of the trigger efficiency curve parameters during run~1 was studied with different measurements, remaining stable with differences of less than 2\% throughout the run.   

The energy was calibrated using the binding energies of the K-shell (10.37~keV) and L-shell (1.30~keV) from the decays of $^{68}$Ge/$^{71}$Ge inside the HPGe diodes, considering a linear behavior in this energy range. At shallow depth, the $^{68}$Ge/$^{71}$Ge radioisotopes are continuously produced by cosmic and muon-induced neutrons~\cite{CONUS:2024lnu}. Thus, it is possible to monitor the stability of the energy scale during the whole measurement of this in situ activation, observing variations below 2\%. A specific $^{252}$Cf irradiation was performed at the end of run~1, collecting in 45\,d of the measurement more than 5000 (700)~events from the K (L) shell in each detector. In this way, an energy calibration uncertainty below 5~eV$_{ee}$ is achieved. The HPGe diode also produces 158\,eV X-rays from the binding energy of the M-shell of the $^{68}$Ge/$^{71}$Ge decays. Using the ratio of the K and L-shells, a total of 100~events per detector was expected after irradiation with $^{252}$Cf. Although an indication of such events is seen, no conclusive signal was yet observed due to the proximity to the noise edge and the lack of statistics. 

\begin{figure}[t]
    \centering
    \includegraphics[width=0.6\textwidth]{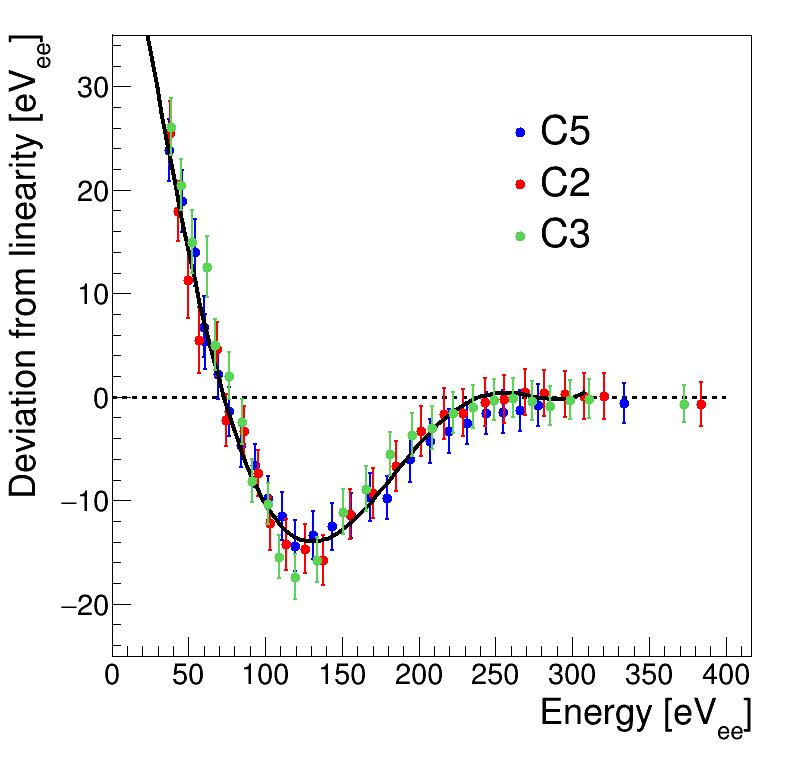}
    \caption{\textbf{Energy non-linearity.} Deviations from a purely linear energy scale measured with a pulse generator for the three \textsc{Conus+} detectors. The trend of the data points are described by the black line. The vertical bars on the data points (colored circles) represent the uncertainties at a 68\% CL (1$\sigma$).
\label{fig:methods_2}}
\end{figure}

The linearity of the DAQ chain in the sub-keV$_{ee}$ region was investigated using the pulse generator signals. The results are shown in Extended Data Fig.~\ref{fig:methods_2}. Deviations from a pure linear behavior are observed below 250~eV$_{ee}$, where a few eV$_{ee}$ variations have a strong impact on the CE$\nu$NS signal. They were attributed to two DAQ related effects~\cite{Bonhomme:2022lcz}. The energy non-linearity was corrected in the CE$\nu$NS analysis and was taken into account during the event energy reconstruction. 

The energy resolution of the detectors at low energies was also evaluated with the pulse generator signals. For the C2 detector, a resolution of ($48\pm1$)\,eV$_{ee}$ (FWHM) was found, while the C5 and C3 detectors have a resolution of ($47\pm1$)\,eV$_{ee}$ (FWHM). An additional contribution to the total energy resolution comes from the statistical fluctuation of the number of electron-hole pairs produced in the Germanium crystals in the case of an event, which is

\begin{equation}\label{eq:resolution}
    \Delta E_S = 2.35 \sqrt{F\epsilon E}
\end{equation}

\noindent with $\epsilon$ = 2.96\,eV, the average energy needed to create a single electron-hole pair in Germanium, and the fano factor $F = 0.1096$. 

\subsection*{Selection cuts}\label{secA1.3}

Selection cuts are applied to reduce background events while keeping the CE$\nu$NS signal. Four different selection cuts are applied to the data: First, the muon-veto system allows for efficient suppression of the impact of cosmic radiation, using a 450~$\mu$s anti-coincidence window between the veto and HPGe signals. The average rate detected in the muon veto during reactor ON is ($274\pm1$)~Hz and decreases to ($214\pm1$)~Hz in reactor off periods. The corresponding average dead times in both periods are 12.3\% (reactor on) and 9.6\% (reactor off). Second, inhibit signals are generated when the increasing baseline has reached saturation of the dynamic range of the transistor reset preamplifier (TRP) for each HPGe detector~\cite{CONUS:2024lnu}. An anti-coincidence window of 1 to 2.5~ms (depending on the detector) is applied to veto unwanted spurious HPGe detector signals generated shortly after the resets. This cut suppressed 30\% events after the muon veto anti-coincidence at low energy, becoming negligible above 5~keV$_{ee}$. The dead time induced by this cut is calculated combined with the previously mentioned muon-veto dead time to avoid the overlapping of both veto windows. An additional dead time of 0.5-2.1\% (depending on the detector) is estimated. Third, the time difference distribution (TDD) of events is studied in each channel as proposed in~\cite{Bonet:2020ntx}. Finally, an anti-coincidence cut is applied between different HPGe detectors with a 5~ms time window. The probability of a neutrino interacting with different detectors is negligible, while for other backgrounds, such as muon-induced neutrons created in the shield, simultaneous hits in several detectors at once can be expected. 

\begin{figure}[t]
    \centering
    \includegraphics[width=\textwidth]{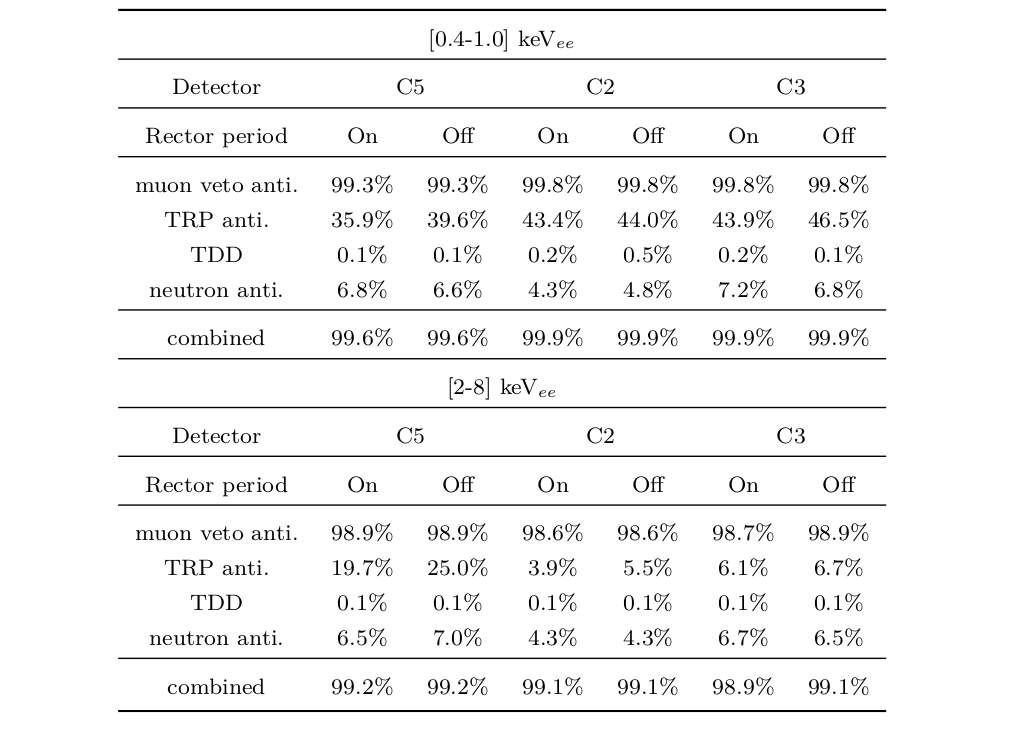}
    \captionof{table}{\textbf{Selection cuts.} The efficiency of all the selection cuts applied during the data processing are listed. The efficiency of each cut is calculated with respect to the previous selection applied. Two energy ranges are considered between [0.4-1]~keV$_{ee}$ and [2-8]~keV$_{ee}$ to avoid the inclusion of background lines and the antineutrino signal. A combined efficiency of over 99\% is achieved for all detectors.
\label{tab_methods1}}
\end{figure}

The rejection efficiencies of these selection cuts are summarized in Extended Data Table~\ref{tab_methods1} for the three detectors in different energy regions. The total dead time induced by the selection cuts is between 12.8\% and 14.4\% in the reactor on periods.

\subsection*{Background model}\label{secA1.4}

The background model used in the analysis of the \textsc{Conus+} data is based on Monte Carlo simulations using the Geant4-based framework MaGe \cite{Boswell:2010mr} following the approach taken in \cite{Bonet:2021wjw}. A complete decomposition of the background in both reactor on and off data was achieved for all detectors used in the analysis. In the following, the main sources of background are described. 

\subsubsection*{Cosmogenic neutrons}\label{secA1.4.1}

As described in \cite{CONUS:2024vyx}, the impact of cosmogenic neutrons with energies up to 100\,MeV was investigated by first propagating the expected neutron spectrum at the KKL location (based on \cite{Goldhagen:2004},\cite{Gordon:2004}) with a flux of $(1.4\pm0.2)\cdot 10^{-2}$~neutrons~s$^{-1}$cm$^{-2}$ through a reactor building model. Then the resulting flux is tracked inside the \textsc{Conus+} room. This flux was used as the basis for a next simulation, in which neutrons were started isotropically from a half-sphere around the \textsc{Conus+} shield and their contribution to the \textsc{Conus+} background was measured. The simulations show a contribution of $(21.6\pm3.1)$~counts~d$^{-1}$kg$^{-1}$ in each detector in the energy range between 0.4~keV$_{ee}$ and 1~keV$_{ee}$, which corresponds to approximately half of the background counts in this region. 

\subsubsection*{Cosmogenic muons}\label{secA1.4.2}

For the muon simulations~\cite{Bonet:2021wjw} the expected flux and muon spectrum at an overburden of 7.4~m~w.e. were calculated from those at Earth's surface (\cite{Bugaev:1998},\cite{Reyna:2006}) and propagated through the shield. The resulting spectrum was validated by comparing it with the \textsc{Conus+} data without the applied muon veto cut. Good agreement was found. The muon veto cut was then applied by multiplying the simulation output by a factor of 0.01 for all energies greater than 2~keV$_{ee}$, corresponding to a muon veto efficiency of 99\%. 

For energies below 2~keV$_{ee}$ a different approach was taken. Here, simulations show an inefficiency in the tagging ability of the muon veto system due to the setup of the shield. The outer muon veto is located under a layer of lead in the \textsc{Conus+} shield. As such, it is possible for muons to pass through this outermost lead layer without hitting one of the muon veto layers. These muons can induce electromagnetic showers in the outer lead layer, which propagate through the shield and are registered in the Germanium detectors. However, because no muon passes through any muon veto layer in such an event, the energy deposition in the plastic scintillator plates will be much lower, resulting in a greatly reduced tagging efficiency of these events. Simulations show that at energies below 0.4~keV$_{ee}$ up to 80\% of all muon-induced background come from such events. Based on this simulation output, this inefficiency was modeled with a polynomial and accounted for in the final muon veto efficiency. The resulting efficiency drops towards lower energies with its minimum being 97\% below 0.4~keV$_{ee}$. Using this approach, the overall background contribution of cosmic ray muons was found to be $(17.4\pm0.3)$~counts~d$^{-1}$kg$^{-1}$ in each detector in the energy range between 0.4~keV$_{ee}$ and 1~keV$_{ee}$, which corresponds to approximately a third of the background counts in this region.

\subsubsection*{Leakage test component}\label{secA1.4.4}

During the final run of the \textsc{Conus} experiment in at KBR an additional background component had to be introduced in the background model~\cite{CONUSCollaboration:2024oks}. This component was present after ventilation of the cryostats with argon gas to avoid mechanical deformation during a regular leakage test at KBR in July 2019~\cite{Bonet:2021wjw}. Of the four detectors used in the \textsc{Conus} experiment, two (C2 and C3) are used for the analysis presented in this work and are affected by this additional background. Indeed, the simulations and the background model show that an additional component with the same shape is still present in the background of these two detectors but is absent in C5 which was not at KBR at the time. This additional background is constant during reactor on and off periods. Therefore, the leakage test component was again included in the background model of C2 and C3 by modeling it using a function with two parameters, as in~\cite{CONUSCollaboration:2024oks}. The resulting impact is below 10\% in the energy region between 0.4\,keV$_{ee}$ and 1\,keV$_{ee}$. 

\subsubsection*{Other background components}\label{secA1.4.3}

The remaining background in each detector is made up of many different components, similar to the situation in KBR~\cite{Bonet:2021wjw}. There are no hints that the detector upgrades or the movement of the setup were introducing any contamination. The background model includes cosmogenically induced isotopes in the Copper parts ($^{57}$Co, $^{60}$Co, $^{54}$Mn) of the cryostat and the Germanium crystals ($^{57}$Co, $^{68}$Ge, $^{68}$Ga, $^{65}$Zn, $^{3}$H), Radon inside the detector chamber, $^{210}$Pb inside the cryostat and shield, metastable germanium states ($^{71m}$Ge, $^{73m}$Ge, $^{75m}$Ge) and inert gases coming from the reactor ($^{85}$Kr, $^{135}$Xe, $^{3}$H). The results of the simulation of these components were scaled to be in accordance with either the rates of gamma lines produced by the decay of these isotopes in the spectrum (e.g.~for radon) or with screening measurements performed prior to the installation. The listed contributions are typically very subdominant in the region of interest, with the decay of radon inside the detector chamber being the only exception. This contribution results in a background rate of $(1.9\pm0.1)$~counts~d$^{-1}$kg$^{-1}$ between 0.4~keV$_{ee}$ and 1~keV$_{ee}$ for the C5 detector (C2: $(2.8\pm0.1)$~counts~d$^{-1}$kg$^{-1}$; C3: $(2.6\pm0.1)$~counts~d$^{-1}$kg$^{-1}$). These values correspond to approximately 5\% of the background in this energy region. Radon decays have a high impact on energies above 100\,keV$_{ee}$, where they can contribute up to 60\% of the measured background. Slow pulses arising from decays on the surface of the diode and in the transition layer are included in the model~\cite{Bonet:2021wjw}. Energy depositions from these events can be stopped within the transition layer and the released charge diffuses slowly into the active volume, resulting in long rise times and incomplete charge collection. Their impact is accounted for by registering the exact coordinates of an interaction in the germanium crystal. If the coordinates place it within the transition layer of the crystal, the energy of the event is shifted towards lower energies using a sigmoid-like function. Details on this procedure can be found in \cite{Bonet:2020ntx}.

\subsubsection*{Model differences in reactor off data}\label{secA1.4.5}

\begin{figure}[t]
    \centering
    \textbf{a}
    \includegraphics[width=0.47\textwidth]{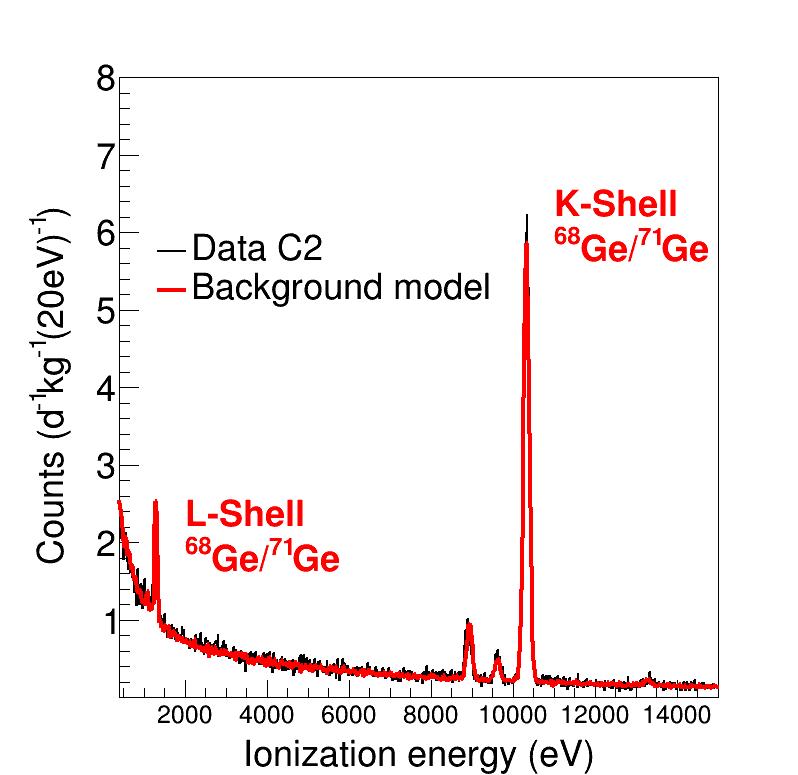}
    \textbf{b}
    \includegraphics[width=0.47\textwidth]{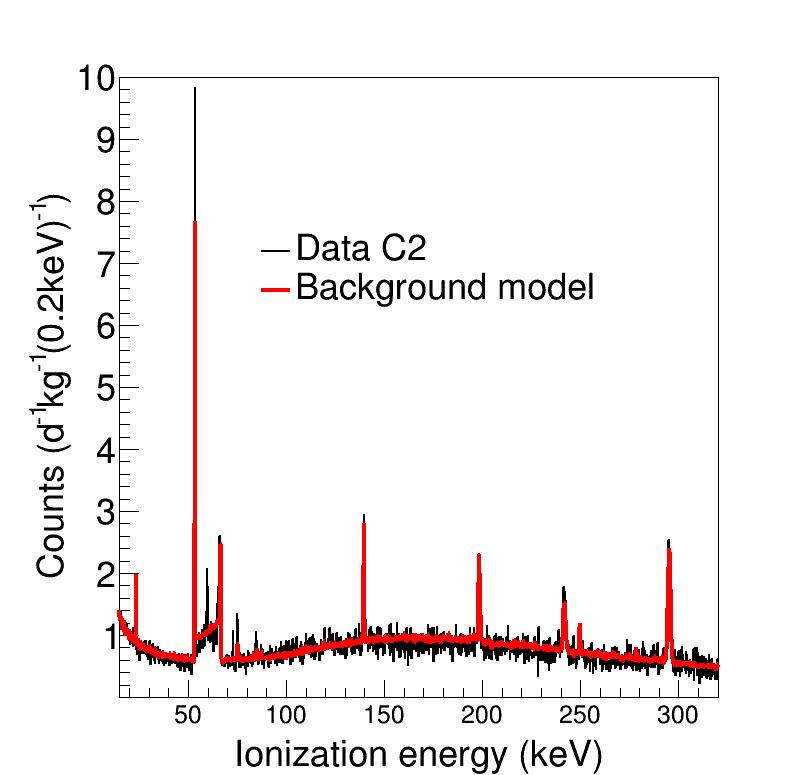}
    \textbf{c}
    \includegraphics[width=0.47\textwidth]{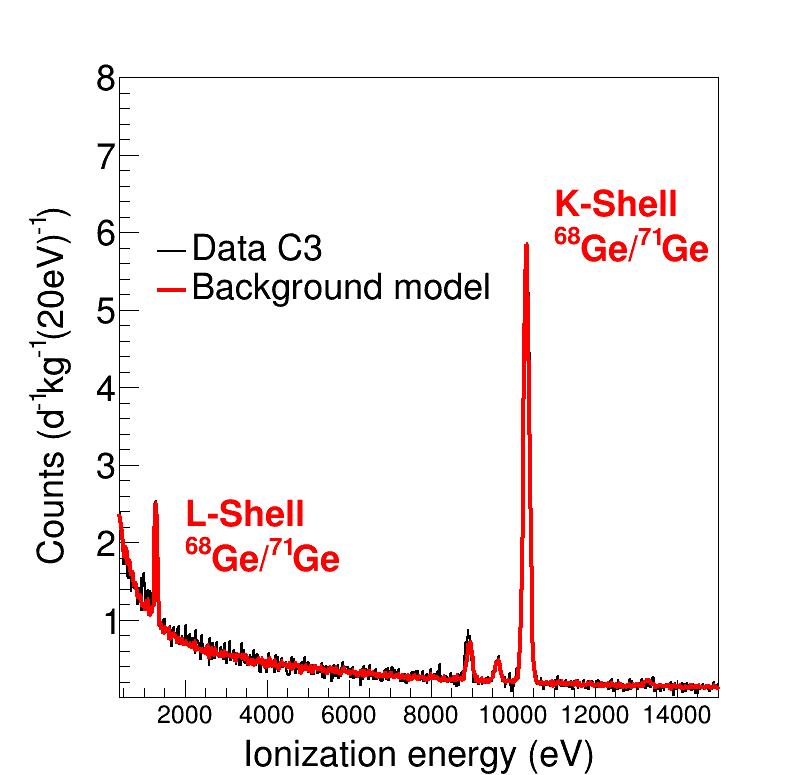}
    \textbf{d}
    \includegraphics[width=0.47\textwidth]{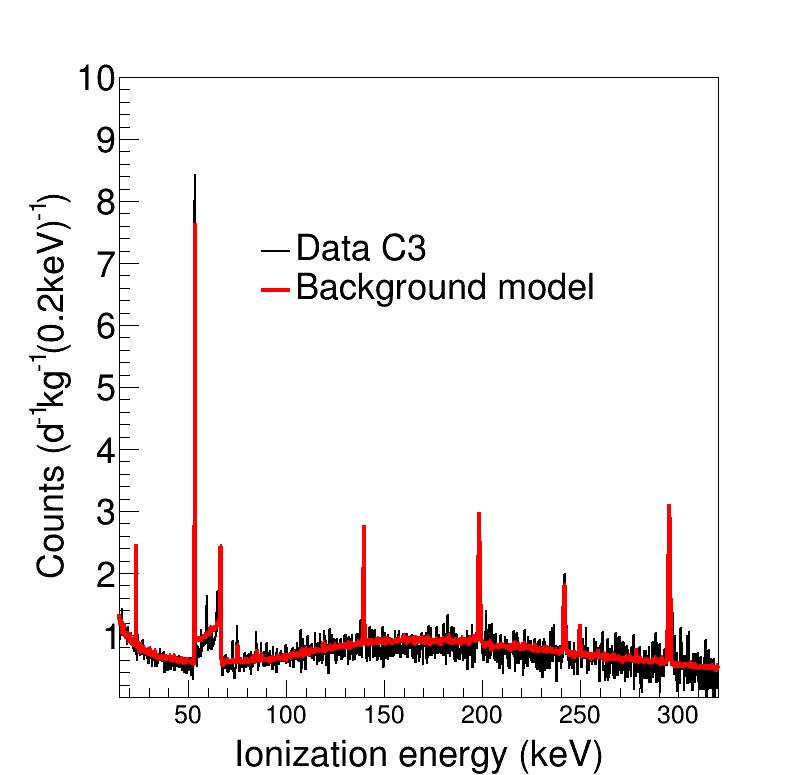}
    \vspace{0.5cm}
    \caption{\textbf{Comparison of data and background model. a) and b)} Data and background model for the C2 and \textbf{c) and d)} C3 detectors, equivalent to figure~\ref{background2}. Good agreement is found between data and model.
\label{methods_bkg_C2C3}}
\end{figure}

The background model accounts for the experimental differences during data collection with the reactor off. The first of these differences is induced by the fact that during a reactor outage, the drywell head from the containment structure surrounding the reactor core is placed directly above the \textsc{Conus+} room. This drywell head is made of 3.8\,cm steel and therefore increases the overburden of the experiment by approximately 0.3\,m~w.e. which results in a reduction of 19\% in the flux of cosmogenic neutrons and a reduction of 3\% in the flux of cosmic ray muons. As a result, these two background contributions are reduced accordingly. The second difference in the background model of the reactor off period comes from a more effective removal of radon in the detector chamber. During the course of run~1 shield tightness and radon free air flushing were improved. As a result, the radon contribution in reactor off time is reduced by approximately a factor of 4 to 6 compared to reactor on. The radon contributions are scaled to match the count rates in the gamma lines induced by the radon decay.  
In addition, reactor-correlated background components, such as reactor neutrons and high energy gammas, e.g.~from $^{16}$N, were investigated. Their impact was found to be negligible in all energy regions, including the region of interest. The resulting background model for all three detectors can be seen in Fig.~\ref{background2} and Extended Data Fig.~\ref{methods_bkg_C2C3}. The time-dependent contributions of the M-shell line of the $^{68}$Ge/$^{71}$Ge decays are included in the background model. The difference between the M-shell contribution in reactor on and off periods was estimated based on the measured K-shell count rates. This difference was found to be below 1\% of the observed neutrino signal. 

\begin{figure}[t]
    \centering
    \textbf{a}
    \includegraphics[width=0.47\textwidth]{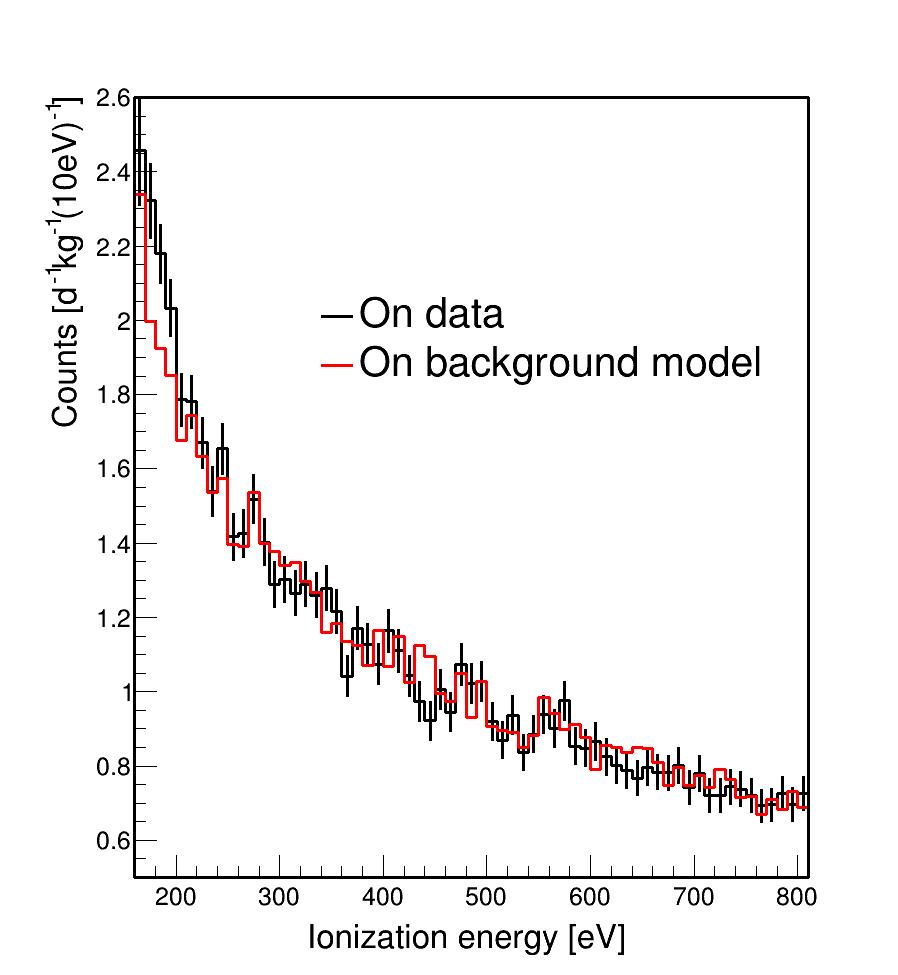}
    \textbf{b}
    \includegraphics[width=0.47\textwidth]{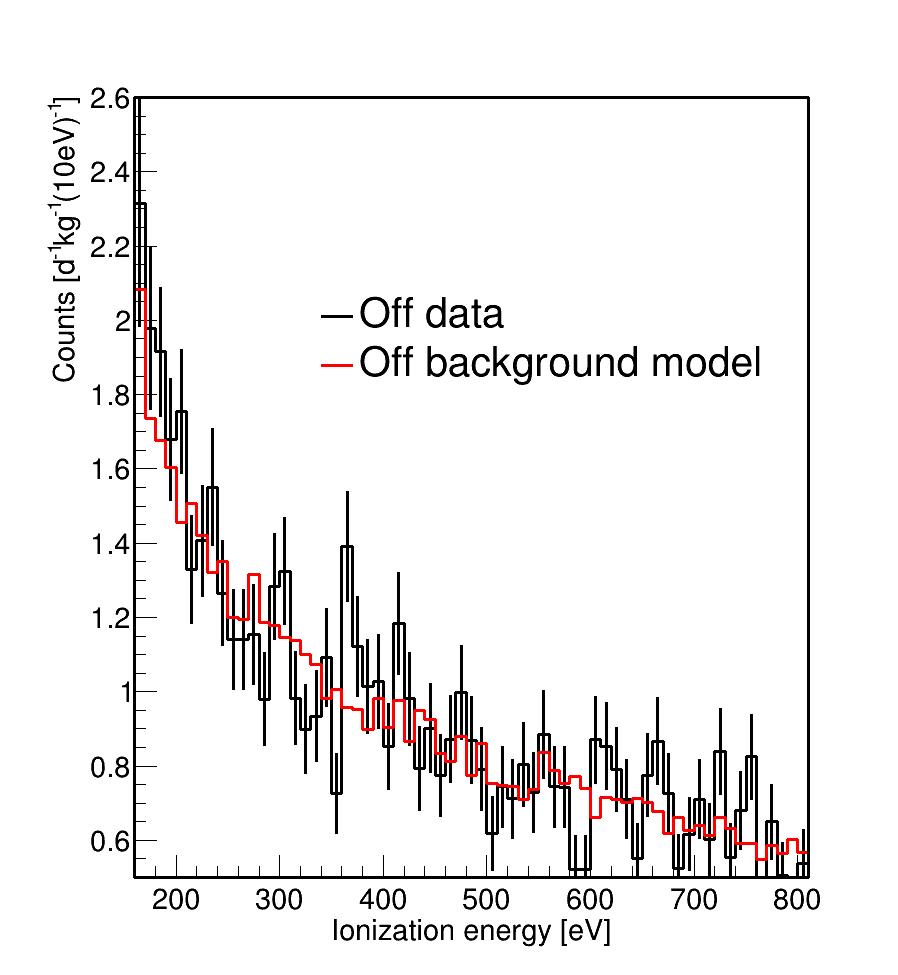}
    \vspace{0.5cm}
    \caption{\textbf{Data versus model in region of interest. a)} Reactor on and \textbf{b)} off count rates normalized to 1\,kg~day in comparison to the corresponding background models are shown in the region of interest. The signal excess in the reactor on data is seen at low energies below 250\,eV$_{ee}$. The vertical bars represent the statistical uncertainties of the data in each 10\,eV$_{ee}$ bin at a 68\% CL (1$\sigma$). Statistical fluctuations are significantly higher in the off than in the on data due to the shorter period of data collection. Due to the slightly different detector thresholds only one detector (C3) contributes to the lowest bin, the second includes two detectors (C3 and C5) and bins above 180\,eV$_{ee}$ are based on the summed spectra of of all three detectors. The difference between the two reactor on curves is shown in figure~3.
\label{methods_bkg_ON-OFF}}
\end{figure}

\begin{figure}[t]
    \centering
    \includegraphics[width=0.9\textwidth]{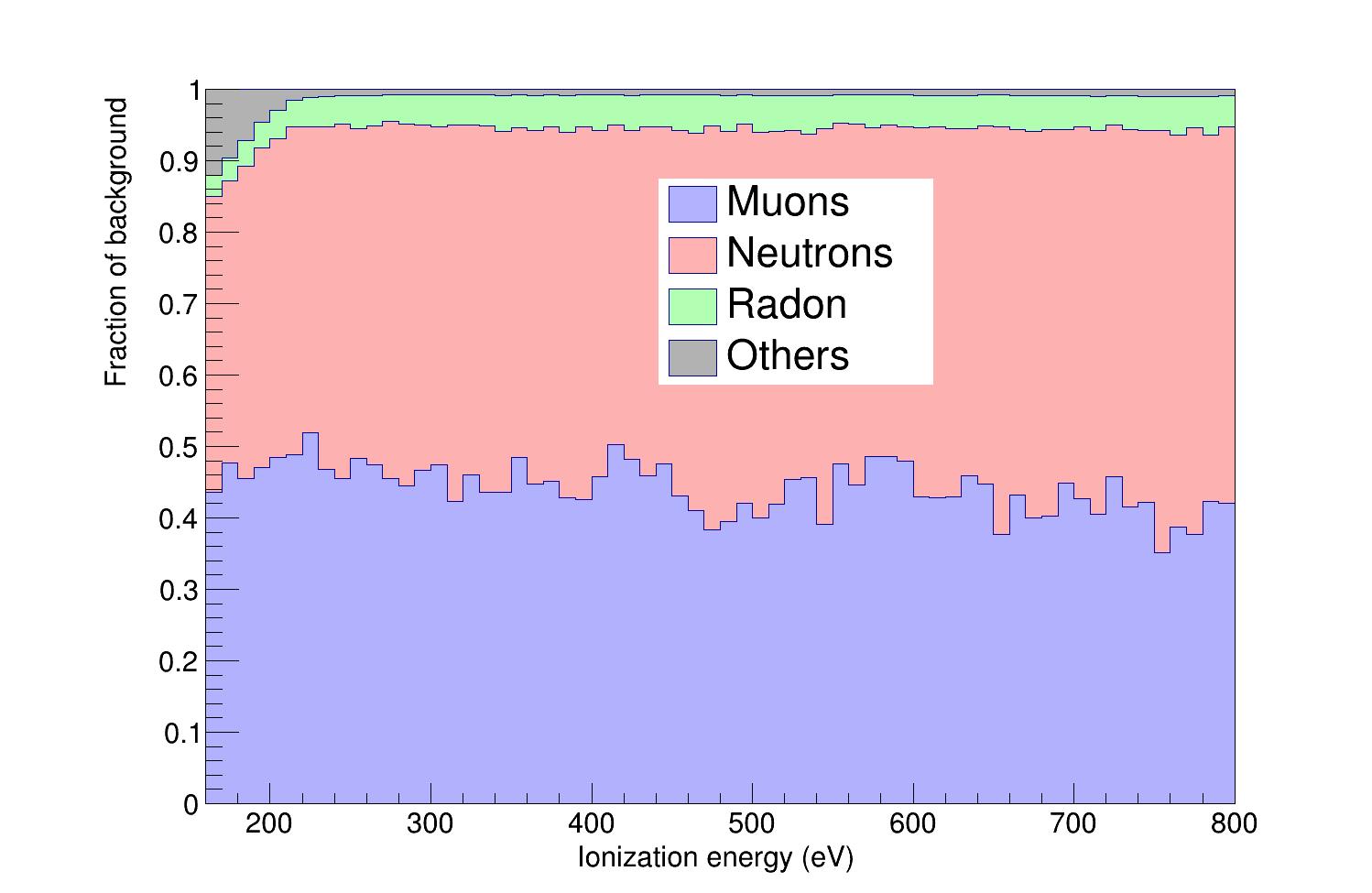}
    \captionof{figure}{\textbf{Fractional background contribution.} The energy dependent fractional contribution of the major background components is shown for one of the CONUS+ detectors (C5). The dominating part are events from cosmogenic origin, neutron and muon-induced background. At energies below 200\,eV$_{ee}$ the M-shell line of the $^{68}$Ge/$^{71}$Ge decays becomes relevant.}
\label{fig:bg_con}
\end{figure}

In Extended Data Fig.~\ref{methods_bkg_ON-OFF}, the reactor on and off data sets are compared with the background models. If the calculated difference between on and off phases is added to the reactor off data, good agreement with the measured on data is found above the signal region. For all detectors background models and data are fully consistent above 0.4\,keV$_{ee}$). Since the data in the reactor off phase is statistically limited, the background model still plays an important role in the analysis. Extended Data Fig.~\ref{fig:bg_con} shows the fractional contribution of the main components to the total background rate as predicted in the model.

\subsection*{Quenching}\label{secA1.5}

The ratio of the ionization energy released by nuclear recoil in a CE$\nu$NS event and the ionization energy of electrons of the same energy is given by the quenching factor. In the \textsc{Conus+} analysis, the energy dependent signal quenching is described by the Lindhard model~\cite{lindhard1963range} with a quenching parameter $k=(0.162\pm0.004)$ as determined in~\cite{Bonhomme:2022lcz}. Alternative quenching descriptions are also tested as described in the supplemental material of~\cite{CONUSCollaboration:2024oks} including a linear and cubic functional form to describe the increased quenching factor compared to Lindhard theory found in~\cite{Collar:2021fcl}. It was shown that the Migdal effect is subdominant with respect to the CE$\nu$NS signal in our region of interest~\cite{AtzoriCorona:2023ais}.

\begin{figure}[t]
    \centering
    \includegraphics[width=0.9\textwidth]{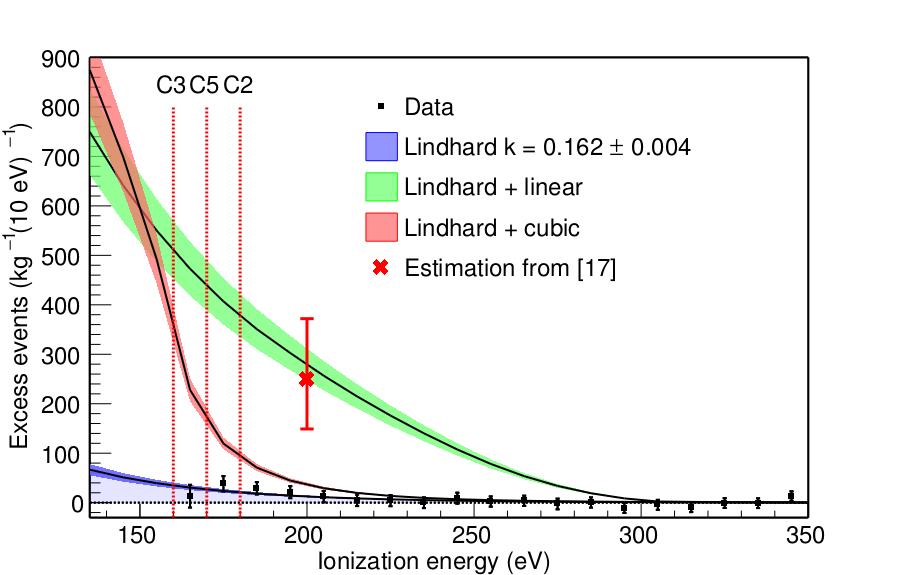}
    \vspace{0.3cm}
    \caption{\textbf{Alternative predictions.} Difference between the data in the reactor on phase and the background model normalized to the total detector mass. The blue bland shows the predicted signal shape assuming the standard Lindhard model with a quenching factor of $k=(0.162\pm0.004)$~\cite{Bonhomme:2022lcz}. The green and red bands depict the signal predictions for the modified Lindhard model based on~\cite{Collar:2021fcl}, using a linear and a cubic function at lower energies respectively. The red vertical lines indicate the energy thresholds of the three detectors used in the analysis. The error bars on the black data points represent uncertainties at a 68\% CL (1$\sigma$). The additional red data point at 200\,eV$_{ee}$ indicates the favored value found in~\cite{Colaresi:2022obx} with a larger error bar reflecting the 2$\sigma$ contour also presented in~\cite{Colaresi:2022obx}.
\label{fig:methods_3}}
\end{figure}

The signal predictions for the different quenching descriptions are shown in Extended Data Fig.~\ref{fig:methods_3} together with the difference between the data in the reactor on phase and the background model. A signal rate of ($2600\pm300$) events is expected for the linear function, while for the cubic function ($550\pm50$) events are predicted. Both numbers are significantly higher than the neutrino rate extracted from the \textsc{Conus+} data. The standard Lindhard model clearly provides the best description of the reactor-correlated excess at low energy. In Extended Data Fig.~\ref{fig:methods_3} we have included an additional data point at 200\,eV$_{ee}$, which is obtained using the information shown in figure 4 of~\cite{Colaresi:2022obx}. There, a CE$\nu$NS signal was approximated with an exponential using two parameters, the amplitude at 200\,eV$_{ee}$ (A$_{0.2}$) and a decay constant. The favored value for A$_{0.2}$ is shown within a 2\,$\sigma$ contour in a 2-dimensional plot of the two parameters. We scaled this A$_{0.2}$ value to our exposure and corrected for the difference in neutrino flux. As expected, it matches the description of Lindhard with a linear increase added at low energies but is in clear conflict with the \textsc{Conus+} data. The lower error bar corresponds to the smallest value of A$_{0.2}$ in the 2\,$\sigma$ contour. This value is also clearly ruled out by the \textsc{Conus+} data points.

\subsection*{Likelihood fit and systematics treatment}
 
A likelihood function is used to determine the CE$\nu$NS signal in the \textsc{Conus+} run~1 data. 

\begin{equation}
    \begin{aligned}
        -2\log\mathcal{L} & = -2\log\mathcal{L}_{\mathrm{ON}} -2\log\mathcal{L}_{\mathrm{OFF}} \\
        & + \sum_{ij} (\bf{\theta}_{i} - \bf{\bar{\theta}}_{i})^T \ \text{Cov}_{ij}^{-1}\  (\bf{\theta}_{j} - \bf{\bar{\theta}}_{j})\\
        & + \sum_{i} \frac{(\theta_{i} - \bar{\theta}_{i})^2}{\sigma_{\theta_{i}}}\, ,
    \end{aligned}
\end{equation}

 $\mathcal{L}_{\mathrm{ON}}$ and $\mathcal{L}_{\mathrm{OFF}}$ are the binned likelihood functions for reactor on and off periods. Gaussian pull terms for the systematic uncertainties are also included. Here, the first term represents the pull terms for correlated parameters, namely the trigger efficiency parameters, and the second term represents pull terms for uncorrelated parameters, like the active mass of the detectors, the reactor neutrino flux and the uncertainty on the energy scale calibration of the spectra. Parameters that were experimentally determined are pulled to their measured values. The background scaling factor $b$, an additional fit parameter for the overall normalization of the background model, is also included and pulled to one.
 The binned likelihood functions assume a Poisson distribution and have the form 
 \begin{equation}
    \begin{aligned}
        -\log\mathcal{L} & = \sum_{i=1}^{n} -n_i \log(\mu_i) + \mu_i + \log(n_i!),
    \end{aligned}
\end{equation}
\noindent where N is the number of bins in the region of interest, $n_i$ is the bin content in the histograms of the measured data and $\mu_i$ is the bin content of the model. For a single detector, the model $\mu$ for reactor on data is calculated from the sum of the background model $n^b$ (scaled with $b$) and the predicted CE$\nu$NS spectrum $n^s$ (scaled with the signal parameter $s$) by taking into account the live time of the experiment ($t_{ON}$ and $t_{OFF}$), the active volume of the detector ($m_{act}$), the dead time correction ($c_{dt}$), the fission flux ($\theta_2$) and a multiplication factor ($\theta_3$), summarizing uncertainties of the detector response. In total: 
\begin{equation}
    \begin{aligned}
        \mu_i^{ON} = ( s \cdot t_{ON} \cdot \theta_2 \cdot n_i^s + b \cdot n_i^b * \frac{t_{ON}}{t_{OFF}}) \cdot c_{dt} \cdot \frac{\theta_3}{m_{act}}, \; \;  
        \mu_i^{OFF} = b \cdot n_i^b \cdot c_{dt} \cdot \frac{\theta_3}{m_{act}}
    \end{aligned}
\end{equation}
The combined fit minimizes $-\log\mathcal{L}_{\mathrm{ON}}$ and $-\log\mathcal{L}_{\mathrm{OFF}}$ for all three detectors simultaneously while the signal parameter $s$, which indicates the number of CE$\nu$NS counts, is shared among detectors.

\begin{figure}[t]
    \centering
    \includegraphics[width=0.8\textwidth]{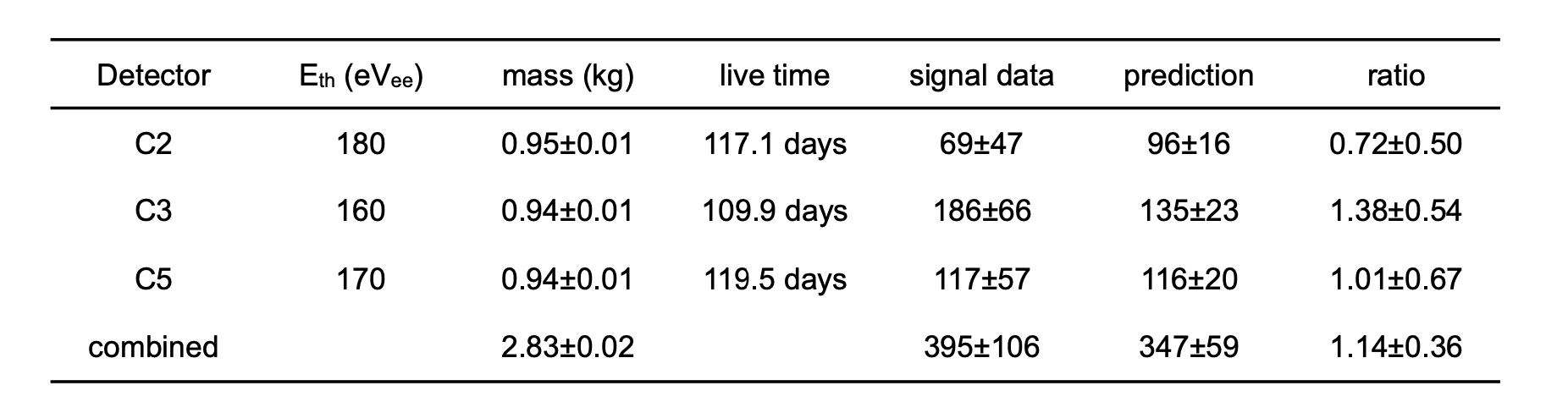}
    \captionof{table}{\textbf{Single detector results.} Energy thresholds ($E_{th}$), reactor on live time and signal events of data and predictions are listed for the individual detectors and the combined fit. In the combined fit result, additional systematic uncertainties mainly from the non-linearity correction and background model are included. The calculated number of neutrino events in the region of interest and the ratio of prediction to data are given for 327~kg\,days of reactor on time.
\label{tab_methods2}}
\end{figure}

Fits using just single detectors independently give consistent results as shown in Extended Data Table~\ref{tab_methods2}. To cross-check the result, fits were performed by two independent likelihoods with different approaches concerning the uncertainty on the quenching model. Likelihood A applied a predicted CE$\nu$NS spectrum with a fixed k parameter ($k=0.162$) while introducing a fourth order polynomial to vary the shape of the signal spectrum applying Gaussian pull terms on each parameter. In likelihood B, the k value of the Lindhard model is a fit parameter with a pull term. Moreover, there are some differences in the treatment of the non-linearity corrections and the minimization algorithms between the two likelihood fits. The results of both likelihood implementations agree within 2\%. 

\begin{figure}[t]
    \centering
    \includegraphics[width=0.6\textwidth]{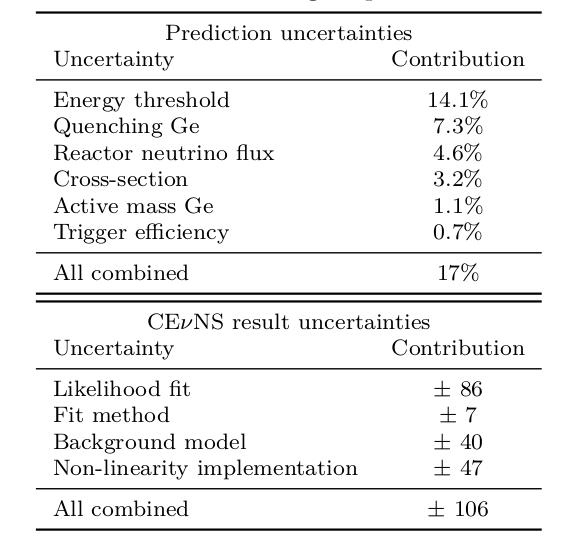}
    \captionof{table}{\textbf{Prediction uncertainties.} Overview of uncertainties on signal prediction and on the CE$\nu$NS result.
    \label{tab_methods3}}
\end{figure}

For the signal prediction, we choose the Helm parameterization of the nuclear form factor~\cite{Engel:1991wq,Bertone:2010zza} and a reactor antineutrino spectrum based on the data-driven approach of~\cite{DayaBay:2021dqj}. The given antineutrino spectra are adjusted to the fission fraction of the Leibstadt reactor and augmented with measured antineutrino spectra at energies above 8\,MeV~\cite{DayaBay:2022eyy} and simulation data below 1.8\,MeV (the threshold of inverse beta decay)~\cite{Estienne:2019ujo}. We assume no uncertainty on the nuclear form factor, but account for a 2\% uncertainty on the weak mixing angle~\cite{Cadeddu:2019eta} leading to an overall uncertainty of 3.2\% on the CE$\nu$NS cross section. The shape of the applied antineutrino spectra, the thermal power of the reactor, fission fractions and energy releases per fission of the relevant fission isotopes contribute a combined 4.6\% error to the expected event rate. Further uncertainties on quenching, detector active mass, trigger efficiency and the energy threshold are taken into account and listed in Extended Data Table~\ref{tab_methods3}. All contributions lead to an overall uncertainty of 17\% on the signal prediction. To be conservative, we assume prediction uncertainties are fully correlated between single detectors. Currently, the dominant uncertainty is from the energy calibration. Future extended calibrations will allow us to reduce this uncertainty from 5\,eV$_{ee}$ to 3\,eV$_{ee}$.

\begin{figure}
    \centering
    \includegraphics[width=\textwidth]{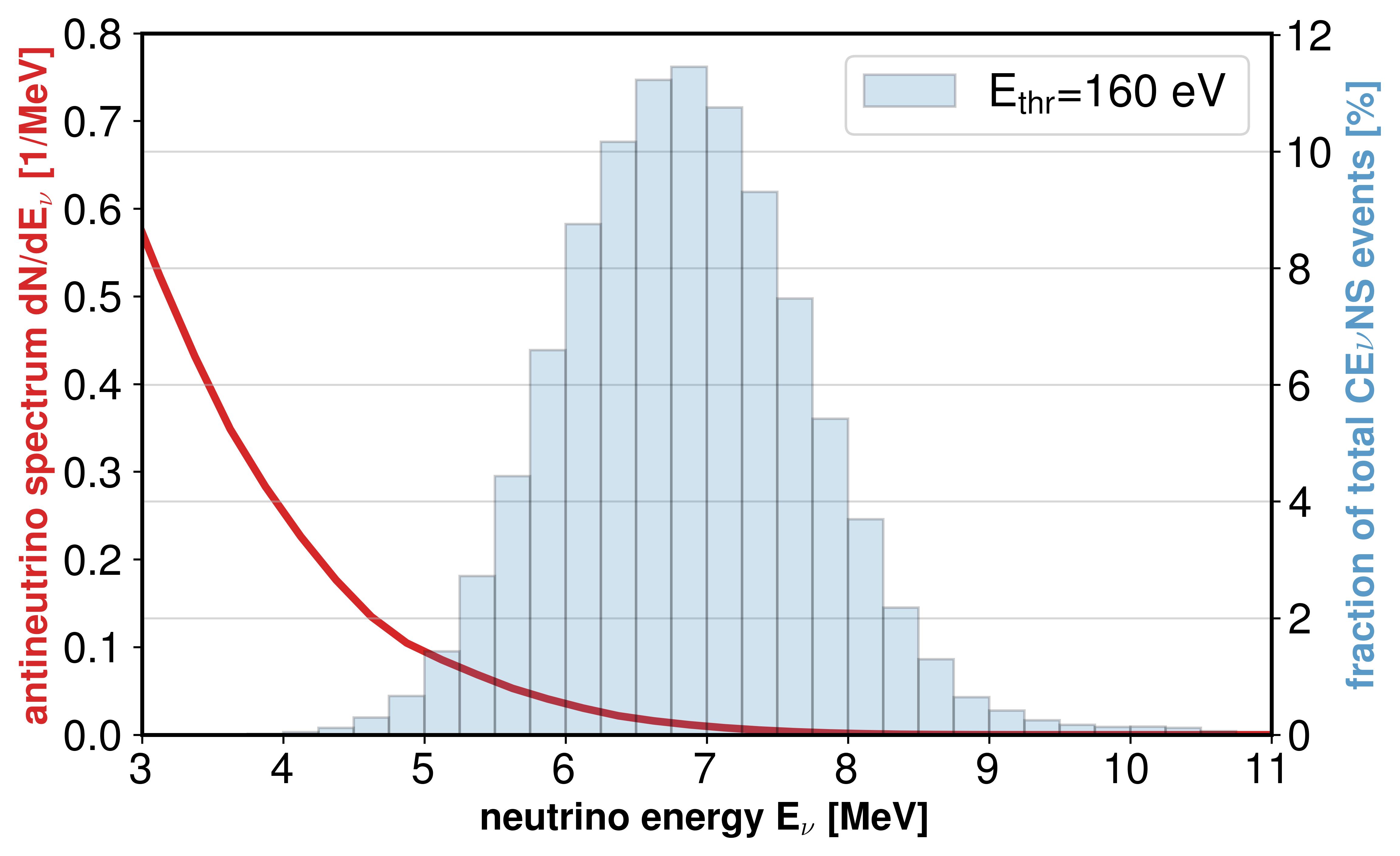}
    \caption{\textbf{Antineutrino spectrum.} The fractional contribution of different energy bins in the antineutrino spectrum to the predicted \textsc{Conus+} signal events is shown for a threshold of 160\,eV$_{ee}$. The red line indicates the antineutrino spectrum above 3\,MeV as emitted from the reactor.
\label{fig:speccon}}
\end{figure}

The likelihood fit itself gives a result of ($395\pm86$) CE$\nu$NS counts in the combined fit of the three detectors. Additional systematic uncertainties not implemented as pull terms are evaluated in a second step and added in quadrature, giving the final uncertainty of $\pm106$~signal counts, as shown in the Extended Data Table~\ref{tab_methods3}. The non-linearity term was obtained by varying the correction parameters and checking the impact on the likelihood result. The uncertainties of the calibration points in Extended Data Fig.~\ref{fig:methods_2} were used to generate the parameter variations. A Gaussian fit was performed on the distribution of the central values of the new likelihood results, where its 1\,$\sigma$ value was taken as additional systematic uncertainty. The term related to the background model was studied in the same way by varying muon flux, neutron flux, and the leakage component. The variations were 14\% (taken from \cite{Goldhagen:2004}), 6\% (based on~\cite{Reyna:2006}) and 10\%, respectively. Additionally, the systematic uncertainty of the fit method was estimated by the difference between likelihood A and B.

The fractional contribution from antineutrinos of different energies to the signal expectation is quantified in Extended Data Fig.~\ref{fig:speccon}. At the current detector threshold, we are sensitive to antineutrino energies above 5\,MeV. With lower detection thresholds, the steep rise in antineutrino flux toward low energies will result in significantly higher signal expectations~\cite{CONUS:2024lnu}.

\bibliography{CONUS}

\end{document}